\documentclass[
  journal=pasa,
  manuscript=research-paper, 
  year=2022,
  volume=40,
]{cup-journal}

\usepackage{microtype,siunitx,booktabs,amssymb}
\title[ORCs as phoenixes of radio galaxies]{Are Odd Radio Circles phoenixes of powerful radio galaxies?}

\author{S. S. Shabala}
\affiliation{School of Natural Sciences, Private Bag 37, University of Tasmania, Hobart 7001, Australia}
\alsoaffiliation{ARC Centre of Excellence for All Sky Astrophysics in 3 Dimensions (ASTRO 3D)}
\email[S. S. Shabala]{stanislav.shabala@utas.edu.au}

\author{P. M. Yates-Jones}
\affiliation{School of Natural Sciences, Private Bag 37, University of Tasmania, Hobart 7001, Australia}
\alsoaffiliation{ARC Centre of Excellence for All Sky Astrophysics in 3 Dimensions (ASTRO 3D)}

\author{L. A. Jerrim}
\affiliation{School of Natural Sciences, Private Bag 37, University of Tasmania, Hobart 7001, Australia}

\author{R. J. Turner}
\affiliation{School of Natural Sciences, Private Bag 37, University of Tasmania, Hobart 7001, Australia}

\author{M. G. H. Krause}
\affiliation{Centre for Astrophysics Research, University of Hertfordshire, College Lane, Hatfield AL10 9AB, United Kingdom}

\author{R. P. Norris}
\affiliation{School of Science, Western Sydney University, Locked Bag 1797, Penrith, NSW 2751, Australia.}
\alsoaffiliation{CSIRO Space and Astronomy, Australia Telescope National Facility, PO Box 76, Epping NSW 1710, Australia.}

\author{B. S. Koribalski}
\affiliation{CSIRO Space and Astronomy, Australia Telescope National Facility, PO Box 76, Epping NSW 1710, Australia.}
\alsoaffiliation{School of Science, Western Sydney University, Locked Bag 1797, Penrith, NSW 2751, Australia.}

\author{M. Filipovi\'c}
\affiliation{School of Science, Western Sydney University, Locked Bag 1797, Penrith, NSW 2751, Australia.}

\author{L. Rudnick}
\affiliation{Minnesota Institute for Astrophysics, University of Minnesota, 116 Church St. SE, Minneapolis, MN 55455, USA.}

\author{C. Power}
\affiliation{International Centre for Radio Astronomy Research, University of Western Australia, 35 Stirling Highway, Crawley, Western Australia 6009, Australia.}
\alsoaffiliation{ARC Centre of Excellence for All Sky Astrophysics in 3 Dimensions (ASTRO 3D)}
 
\author{R. M. Crocker}
\affiliation{Research School of Astronomy and Astrophysics, Australian National University, Canberra 2611, A.C.T., Australia.}

\doi{10.1017/pasa.2020.32}

\received {dd Mmm YYYY}
\revised  {dd Mmm YYYY}
\accepted {dd Mmm YYYY}
\published{dd Mmm YYYY}


\graphicspath{{}}

\begin{document}

\begin{abstract}

Odd Radio Circles (ORCs) are a class of low surface brightness, circular objects approximately one arcminute in diameter. ORCs were recently discovered in the Australian Square Kilometre Array Pathfinder (ASKAP) data, and subsequently confirmed with follow-up observations on other instruments, yet their origins remain uncertain. In this paper, we suggest that ORCs could be remnant lobes of powerful radio galaxies, re-energised by the passage of a shock. Using relativistic hydrodynamic simulations with synchrotron emission calculated in post-processing, we show that buoyant evolution of remnant radio lobes is alone too slow to produce the observed ORC morphology. However, the passage of a shock can produce both filled and edge-brightnened  ORC-like morphologies for a wide variety of shock and observing orientations. Circular ORCs are predicted to have host galaxies near the geometric centre of the radio emission, consistent with observations of these objects. Significantly offset hosts are possible for elliptical ORCs, potentially causing challenges for accurate host galaxy identification. Observed ORC number counts are broadly consistent with a paradigm in which moderately powerful radio galaxies are their progenitors.

\end{abstract}

\section{Introduction}
\label{sec:intro}

Odd Radio Circles (ORCs) are circles of low surface brightness radio continuum emission, first discovered in the Australian SKA Pathfinder (ASKAP) telescope Evolutionary Map of the Universe (EMU) pilot survey data (Norris et al. 2020). Eight of these rings or edge-brightened disks have so far been found in the 800-1088 MHz ASKAP data \citep{NorrisEA2021_orcs, NorrisEA2022, KoribalskiEA2021, FilipovicEA2022, GuptaEA2022}. ORCs are characterised by low surface brightness ($100-300\, \mu$Jy/beam; ASKAP rms sensitivity is $\sim 30\, \mu$Jy/beam), edge-brightened rings approximately $60-80$\,arcsec in diameter. Several ORC detections have been confirmed at both longer wavelengths (325 MHz continuum with the Giant Meterwave Radio Telescope, GMRT) and higher resolution (MeerKAT); and discovered with other instruments \citep{LochnerEA2023, KoribalskiEA2023}. The best-studied ORC, dubbed ORC1 \citep{NorrisEA2021_orcs, NorrisEA2022} shows a narrow ring of emission unresolved by the ASKAP $11'' \times 13''$ FWHM beam; the ring is marginally resolved by MeerKAT's $6''$ beam \citep{NorrisEA2022}. It shows a remarkably uniform spectral index \footnote{We adopt the convention $S_\nu \propto \nu^{-\alpha}$ throughout the paper.} $\alpha \sim 1.1$ between 800 and 1400 MHz, with hints of filamentary structure across the ring \citep{NorrisEA2022}.

No electromagnetic counterparts to ORCs have so far been found at other wavelengths. Four ``single ORCs'' (ORC1 and ORC4, \citealt{NorrisEA2021_orcs}; ORC5, \citealt{KoribalskiEA2021}; SAURON, \citealt{LochnerEA2023}) have candidate elliptical host galaxies co-located in projection at the ORC's geometric centre, with photometric redshifts in the range $0.27-0.55$; both ORC1 and ORC5 also have galaxies coincident in projection with the ring structure. ORCs 2 and 3, on the other hand, don't appear to have a central candidate host galaxy; however these two ORCs are in close proximity to each other on the sky, and could be part of the same structure \citep{NorrisEA2021_orcs}. \citet{KoribalskiEA2023} have recently reported discovery of a single ORC without a clear central host in MeerKAT data. In addition to single ORCs and ORCs with companion lobes, several ORC candidates have also been found \citep{GuptaEA2022}.

The origin of ORCs is, at present, a mystery. Several hypotheses have been put forward to explain these, including jet-inflated lobes, black hole mergers \citep{NorrisEA2022}, starburst-driven shocks \citep{CoilEA2023}, tidal disruption events \citep{OmarEA2022}, precessing AGN jets \citep{NoltingEA2023}, merger shocks \citep{DolagEA2023}, and even supernova remnants \citep{FilipovicEA2022}. Indeed, there may be more than one explanation for this morphological class. If ORCs are extragalactic at redshifts suggested by their candidate host galaxies, a very large injection of energy is required to inflate ORCs to their implied sizes of several hundred kiloparsecs. \citet{DolagEA2023} recently presented a detailed numerical model suggesting that shock acceleration from galaxy -- galaxy mergers can produce radio sizes and morphologies similar to the observed ORCs. Supermassive black holes at galaxy centres are another obvious candidate for providing this large amount of energy. The association of radio galaxy lobes with some ORCs suggests that jets may play a role in this process.

Relativistic jets emanating from Active Galactic Nuclei (AGN) at galaxy centres are a key component in regulating the baryon cycle within and outside galaxies. These jets are found in systems with short cooling times \citep{MittalEA2010}, and estimates of their energetics suggest a balance between gas cooling and jet heating \citep{BestEA2005, KaiserBest2007, ShabalaEA2008, TurnerShabala2015, HardcastleEA2019, KondapallyEA2023}. This ``maintenance'' mode of AGN feedback keeps circumgalactic and intracluster gas -- which would otherwise cool rapidly -- hot, and explains the largely quiescent star formation histories of massive ellipticals over the past several Gyr \citep{CrotonEA2006, ShabalaAlexander2009, FanidakisEA2011, RaoufEA2017, WeinbergerEA2017, ThomasEA2021}. In this ``thermostat'' paradigm of jet feedback, the jet duty cycle is in part determined by jet feedback \citep{KaiserBest2007, PopeEA2012}. Observations of radio galaxies with multiple pairs of lobes \citep[so-called Double-Double Radio Galaxies, e.g.][]{SchoenmakersEA2000, SteenbruggeEA2010, KonarHardcastle2013, MahatmaEA2018, JurlinEA2020} provide dramatic observational evidence of jet intermittency; \citet{BruniEA2019} recently showed that a high fraction of Giant Radio Galaxies may show intermittent jet activity. Recent observations of populations of active, remnant and re-started radio jets \citep{JurlinEA2020} and detailed modelling of their dynamics and synchrotron emission \citep{ShabalaEA2020} suggest that Chaotic Cold Accretion \citep{GaspariEA2013, McKinleyEA2022} is a mechanism which can naturally facilitate intermittent black hole accretion and jet activity. Remnant radio lobes are therefore expected to be ubiquitous, especially in environments with a high jet duty cycle such as cool core clusters. Re-acceleration of cosmic ray electrons within the remnant lobes is thought to be responsible for much diffuse radio emission in clusters, including radio relics and radio haloes (see the \citealt{BrunettiJones2014} review and references therein). In this paper, we examine whether remnant AGN lobes are plausible progenitors of ORCs.

\citet[][see also \citealt{BrueggenKaiser2002}]{ChurazovEA2001} modelled the morphological evolution of remnant radio lobes rising buoyantly through a cluster atmosphere. These authors showed that the remnant lobes undergo significant morphological evolution in the buoyant phase: ambient gas is uplifted by the radio lobes through the central channel, and subsequent adiabatic expansion pushes the remnant plasma away from the axis of symmetry; this causes a characteristic ``mushroom'' shape, followed eventually by a torus. Viewed close to end-on, such a torus would produce a ring morphology similar to an ORC.

This scenario, however, cannot explain the existence of ORCs: as we show in Section~\ref{sec:remnantDynamics} (see also \citealt{KaiserCotter2002, GodfreyEA2017, Turner2018, Hardcastle2018a, YatesEA2018, EnglishEA2019, ShabalaEA2020}), once the jets cease to supply energy to the lobes, remnant lobes fade extremely quickly. This situation is exacerbated for extremely large lobes such as those implied by ORCs at non-negligible redshifts, due to the unavoidable inverse Compton losses. As pointed out by \citet{ONeillEA2019} and \citet{NoltingEA2019}, the dynamical behaviour of jet-inflated remnant lobes is more complex than that of bubbles in a cluster atmosphere: remnant lobes inflated by powerful jets will spend a considerable amount of time expanding supersonically through the ambient gas (and rapidly fading) before transitioning to the buoyant phase. Hence the remnant lobes will fade below any realistic detection limit much faster than the dynamical time required to transform dynamically into a torus.

\citet{EnsslinBrueggen2002} showed that passage of a shock through a fossil radio bubble, placed ``by hand'' into the simulation, can produce radio emission with a toroidal morphology. This important result can be understood as follows. As the shock propagates through the cluster gas, the post-shock thermal pressure is balanced by the ram pressure (and a small component of thermal pressure) in the pre-shock gas. As the shock first makes contact with the radio bubble, however, the post-shock thermal pressure now exceeds the pre-shock ram pressure (plus thermal pressure) in the underdense bubble, facilitating rapid expansion through the bubble and formation of a torus. \citet{ONeillEA2019} and \citet{NoltingEA2019} extended this pioneering analysis to more realistic, jet-inflated remnants, and confirmed that toroidal structures can form in these situations. \citet{ZuHoneEA2021} confirmed that the cosmic ray electrons will form toroidal structures when compressed by a shock in a more realistic cluster merger scenario.

While this is promising, the large sizes of ORCs, if these are extragalactic, pose several challenges to the torus hypothesis. To reach transverse sizes of several hundred kpc, ORCs must be inflated by radio sources with similarly large sizes at switch-off. To remain dynamically stable, such lobes can only be inflated by powerful radio sources, which are relatively rare (see Section~\ref{sec:numberCounts}). In this scenario, buoyant bubble models are not applicable, and dynamics of the remnant lobes must be taken into account. This has been done by \citet{NoltingEA2019}, who showed that recently switched off lobes revived by a cluster shock passage can produce toroidal structures $\sim 200$\,kpc in diameter; however those authors only considered recently switched off lobes, which will be a relatively small subset of all shocked remnants. \citet{DolagEA2023} showed that ORC-like structures can be produced by shocks resulting from galaxy mergers, but pointed out that the energetics required to produce the observed radio emission through diffusive shock acceleration of thermal electrons are challenging.

In this paper, we explore scenarios under which shock acceleration of remnant radio lobes can produce ORC-like radio morphologies. Our relativistic hydrodynamic simulations follow the full jet duty cycle, from inflation of supersonically expanding lobes by the initially conical, relativistic jets, through the remnant phase of lobe evolution, to shock re-acceleration of the lobe plasma. At all stages, we calculate in post-processing the synthetic radio emission from lobe electrons; this includes re-acceleration at shocks, as well as adiabatic, synchrotron and inverse Compton losses. This approach allows us to self-consistently follow the populations of cosmic-ray electrons available for re-acceleration by the shock passage.

We introduce our technical setup and simulations in Section~\ref{sec:simulations}. We present our results in Sections~\ref{sec:results} and \ref{sec:phoenices}. In Section~\ref{sec:discussion} we discuss our main findings, focusing on the importance of progenitor properties, remnant age, and shock and viewing geometry. We summarise in Section~\ref{sec:summary}.
 
\section{Methods}
\label{sec:simulations}

\subsection{Numerical hydrodynamics}
\label{sec:technicalSetup}

We use the freely-available numerical hydrodynamics code PLUTO\footnote{http://plutocode.ph.unito.it} version 4.3 \citep{MignoneEA2007, MignoneEA2012} to simulate the evolution of initially relativistic bipolar AGN jets. Our setup follows \citet{YatesEA2018} and \citet{YatesJonesEA2021, YatesJonesEA2022}, and we refer the reader to these papers for technical details. Briefly, we use the relativistic hydrodynamics physics module of PLUTO, along with the {\texttt hllc} Riemann solver, linear reconstruction, second-order Runge-Kutta time-stepping, and a Courant-Friedrichs-Lewy (CFL) number of 0.33. 

\subsubsection{Simulation grid}

\begin{figure}
    \centering
    \includegraphics[width=1.0\textwidth]{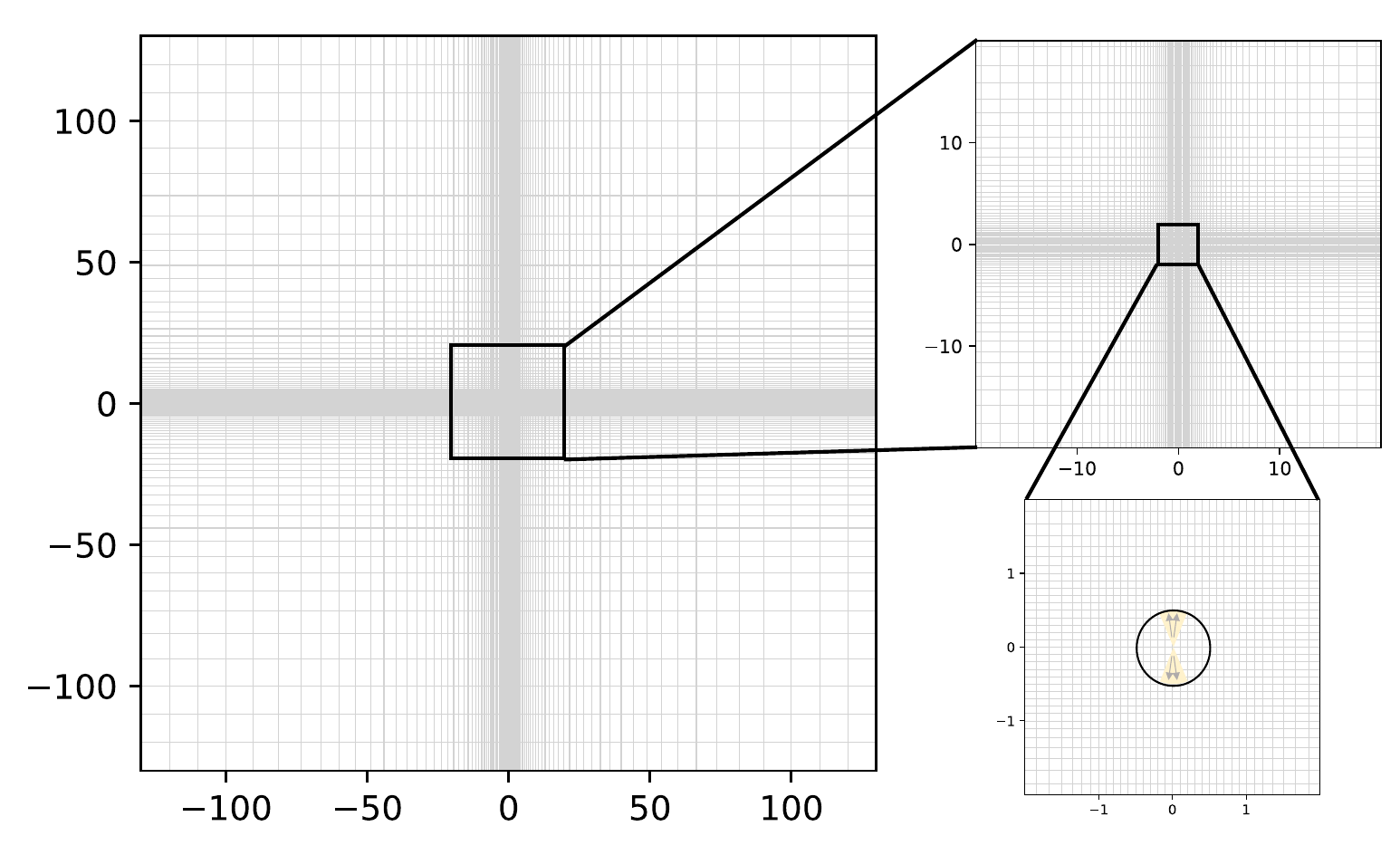}
    \caption{Two-dimensional projection of the simulation grid. Coordinates are in kpc. The central 1\,kpc regions in each coordinate have resolution of 0.1 kpc, decreasing to 1.0\,kpc at a distance of 10\,kpc from the origin, and 10\,kpc resolution at distances beyond 100\,kpc. The jet injection cone and associated spherical region are also shown.}
    \label{fig:grid}
\end{figure}

A challenge in accurately representing large radio sources inflated by relativistic jets is the need to achieve sufficiently high resolution in the jet collimation region while also simulating a large ($>1$\,Mpc$^3$) simulation grid. Insufficient resolution at jet collimation scale would result in an underexpanded, heavy jet with large forward ram pressure; such a jet will inflate unrealistically narrow radio lobes. We use a static three-dimensional Cartesian grid, typically $290^3$ cells consisting of five uniform and stretched patches symmetric about the origin (see Figure~\ref{fig:grid} and Table~\ref{tab:sims}), to achieve this. A central uniform grid patch of 20 cells is defined around the injection region in all three dimensions ($-1 \rightarrow 1$\,kpc, a resolution of 0.1\,kpc). Either side of this central patch is a geometrically stretched grid of 45 cells, spanning $1 \rightarrow 100$\,kpc; this patch has resolution of 1.0\,kpc at 10\,kpc from the origin, and 10.0\,kpc resolution at 100\,kpc. We use a uniform grid of a further 90 cells to maintain this 10\,kpc resolution in the outermost regions ($100 \rightarrow 1000$\,kpc) of the simulation domain. Following Krause et al. (2012), a jet with kinetic power $Q_{\rm j}$, half-opening angle $\theta_{\rm j}$ and speed $v_{\rm j}$ in an external environment with density $\rho_{\rm x}$, sound speed $c_{\rm x}$ and adiabatic index $\Gamma_{\rm x}$ will begin recollimation at a length scale $L_{\rm 1,a} = \left( \frac{\Gamma_{\rm x} \sin \theta}{\pi (1-\cos \theta)} \right)^{1/2} \left( \frac{Q_{\rm j}}{\rho_{\rm x} v_{\rm j}} \right)^{1/2} c_{\rm x}^{-1}$. For our typical parameters (Section~\ref{sec:env}), the expected scale of jet recollimation is $L_{\rm 1,a} \sim 5$\,kpc, and the jet width of 3\,kpc is sufficiently resolved with our 0.5\,kpc resolution at this distance.

\subsubsection{Jet injection}
\label{sec:jet_injection}

Following \citet{YatesJonesEA2021}, the bipolar jets are injected as a mass outflow internal boundary condition. We adopt a half-opening angle of $\theta_{\rm j} = 15^\circ$, which produces Fanaroff-Riley type II \citep[FR-II]{FanaroffRiley1974} jets with realistic jet widths \citep{KrauseEA2012, YatesEA2018, YatesJonesEA2021}. The jet injection region is defined as a sphere with radius $0.5$\,kpc. The values of density and pressure in cells within this region are overwritten with the corresponding injection zone values; cells at angles $\theta < \theta_{\rm j}$ to the $z$-axis within the injection sphere are also assigned a velocity equal to the jet velocity $v_{\rm j}=0.95c$. The jet density $\rho_{\rm j}$, pressure $P_{\rm j}$ and cross-section $A_{\rm j}$ at the jet inlet can be related to the (single-) jet kinetic power \citep{MukherjeeEA2020, YatesJonesEA2021},

\begin{equation}
Q_{\rm j} = \left[ \gamma_{\rm j} (\gamma_{\rm j}-1) c^2 \rho_{\rm j} + \gamma^2 \frac{\Gamma_{\rm j}}{\Gamma_{\rm j}-1} P_{\rm j}  \right] v_{\rm j} A_{\rm j}
\label{eqn:jet_injection}
\end{equation}
where $\gamma_{\rm j} = \left[ 1-(v_{\rm j}/c)^2 \right]^{-1/2}$ is the bulk jet Lorentz factor and $\Gamma_{\rm j}$ is the jet adiabatic index. We inject pressure-matched jets, i.e. $P_{\rm j}$ equals the ambient pressure at the jet inlet. The requirement that the jets are cold ($\chi \equiv \frac{\Gamma_{\rm j}}{\Gamma_{\rm j}-1} \frac{\rho_{\rm j} c^2}{P_{\rm j}} = 100$) allows the jet density at injection to be calculated. Although $\Gamma_{\rm j}=5/3$ (corresponding to cold jets) initially, we use the Taub-Mathews equation of state \citep{Taub1948, Mathews1971, MignoneMcKinney2007} to account for any shock-heating of the jet material; the reader is referred to \citet{YatesJonesEA2021} for further details of the jet injection setup.

\subsubsection{Environment}
\label{sec:env}

For all simulations, we adopt an environment representative of clusters. We adopt an isothermal beta profile for the density and pressure, $\rho_{\rm x}(r) = \rho_0 \left[1 + \left( \frac{r}{r_0} \right)^2 \right]^{-3\beta}$, with temperature $T=3.6 \times 10^7$\,K (corresponding to a sound speed $c_s = 910$\,km\,s$^{-1}$), central density $\rho_0=5 \times 10^{-23}$\,kg\,m$^{-3}$, core radius $r_0=30$\,kpc, and exponent $\beta=0.38$. These values are consistent with observed low-redshift clusters \citep{VikhlininEA2006} and cosmological simulations \citep{CuiEA2018, YatesJonesEA2023}, with the possible exception of a smaller scaling radius $r_0$; as most of the evolution presented in this paper occurs on scales of several hundred kpc, our environments are representative of clusters on the scales of interest.

This choice of a cluster environment is consistent with observations of both ORCs and their putative progenitors,  powerful radio galaxies. Probable host galaxies for ORC1, ORC4 and ORC5 are all massive, red ellipticals with slowly accreting black holes \citep{KoribalskiEA2023,RupkeEA2023}. When hosting powerful FR-II radio sources, such galaxies are preferentially found in clusters \citep{HardcastleCroston2020}. While ORC environments are not well constrained at present, ORC1 is likely to be in an overdensity and likely hosts of ORCs 4 and 5 have close companion galaxies \citep{NorrisEA2021_orcEnv}, similar to hosts of powerful radio galaxies \citep{KrauseEA2019}.

We note that our analysis below is not strongly affected by the choice of environment. Edge-brightened FR-II radio galaxies such as those simulated in this work are even more prevalent in less dense environments such as galaxy groups \citep{HardcastleCroston2020}, hence lobe dynamics would not be qualitatively impacted. For fixed jet power, the time to inflate radio lobes to a given size (and thus the total number of radiating leptons) scales with density as $\rho^{-1/3}$ \citep{KaiserAlexander1997,ShabalaGodfrey2013,TurnerShabala2023}, hence the luminosities presented in Section~\ref{sec:results} are robust for a realistic range of environments.

\subsection{Radio emission}
\label{sec:radioEmission}

Strong internal shocks such as jet recollimation shocks and hotspots are the sites of particle acceleration in radio galaxy jets \citep{MeisenheimerEA1989, OrientiEA2010, McKeanEA2016}, and strong external shocks have been argued to re-accelerate relic non-thermal plasma \citep{FinoguenovEA2010, IapichinoBrueggen2012, StroeEA2014}.

We follow the method of Yates-Jones et al. (2022), which employs the Lagrangian passive tracer particle module of PLUTO 4.3 \citep{VaidyaEA2018}. The details of our implementation are given in \ref{sec:AppendixRadioEmission}. Briefly, each Lagrangian particle in the PLUTO simulations represents an ensemble of electrons; our simulations track the re-acceleration history due to shocks of each such ensemble, and we calculate in post-processing the energy losses due to adiabatic expansion, inverse Compton upscattering of Cosmic Microwave Background (CMB) photons, and synchrotron emission.

The radio emissivity is calculated as follows. Starting with the Lorentz factor required for the given electron ensemble to radiate at the observed frequency, we iterate backwards in time to infer the (higher) Lorentz factor at the time when this ensemble was last accelerated. This injection Lorentz factor depends on the local magnetic field strength (for synchrotron losses) and the Cosmic Microwave Background energy density (for inverse Compton CMB losses). Electron populations which have suffered severe losses (e.g. in regions of high magnetic field strength and/or accelerated sufficiently long ago) will require very high injection Lorentz factors; such electrons are in the power-law tail of the Diffusive Shock Acceleration (DSA) energy distribution, and will therefore contribute little to the integrated emissivity.

Several model parameters affect the predicted synchrotron emissivity; the majority of these are robustly constrained by observations of radio galaxies and remnants.

The power-law slope of the DSA electron energy distribution in the active phase is set to $s=2.2$, corresponding to a synchrotron spectral index of $\alpha_{\rm inj}=0.6$ characteristic of radio galaxy lobes; we note that the exact value of this parameter is not important to the results presented in this paper.

The greatest source of uncertainty in our analysis is the low-energy cutoff for the radiating particles. We set this parameter to $\gamma_{\rm min}=500$, consistent with observations of hotspots \citep{CarilliEA1991, StawarzEA2007, GodfreyEA2009, McKeanEA2016} and lobe evolutionary tracks \citep{TurnerEA2018a, TurnerEA2018b, YatesJonesEA2022} in powerful radio sources similar to those simulated here. No radio relics have been observed to show a low-energy cutoff; however the much lower (by about three orders of magnitude compared to hotspots; \citealt{GodfreyEA2009}) magnetic fields in remnants mean such a turnover would only be detectable at frequencies below 100 MHz even for high $\gamma_{\rm min}$ values. We therefore adopt a fiducial value of $\gamma_{\rm min} = 500$ in our analysis, but caution that the uncertainty in this parameter introduces a large uncertainty in the normalisation of calculated emissivities, as for spectra steeper than $s=2$ the majority of electrons will have Lorentz factors just above $\gamma_{\rm min}$.

Because the simulations presented in this paper are purely hydrodynamic, we need to assume a mapping between the lobe magnetic field strength and a hydrodynamic quantity. We follow an established approach \citep{KaiserEA1997, TurnerShabala2015, Hardcastle2018a} to relate the lobe magnetic field strength to pressure; this approach yields inferred lobe magnetic field strengths which are consistent with independent X-ray measurements \citep{TurnerEA2018b, InesonEA2017}. Similarly, our assumed remnant magnetic field strengths are consistent with observations; details are provided in \ref{sec:AppendixRadioEmission}. While the lack of magnetic fields in our simulations means that we are unable to track the small-scale details in emergent radio structures, the success of hydrodynamic approaches in modelling radio galaxy lobes validates our overall analysis.

The final important parameter is the total number of radiating particles. Because we simulate the full duty cycle of jet activity, we track this quantity throughout the active phase as jet material is supplied to the lobes, and then ensure that it is conserved in the remnant phase of lobe evolution.

For each simulation snapshot, surface brightness is calculated by integrating emissivity through the entire simulation volume along a given line-of-sight. These synthetic surface brightness maps are convolved with a two-dimensional Gaussian beam with 6 arcsec Full Width at Half Maximum. Redshift dependence is explicitly included in post-processing through changes to the rest-frame frequency (and hence Lorentz factor) of emission for a given observing frequency, strength of the CMB photon energy density field, resolution, and observed flux density of simulated sources.

\subsection{Simulations}
\label{sec:sims}

The suite of simulations used in this work is presented in Table~\ref{tab:sims}. The main simulations use relativistic ($v_{\rm j} = 0.95c$) jets with kinetic power $10^{38}$\,W per jet, powered for 50\,Myr; these parameters are characteristic of moderate power FR-II radio galaxies \citep{RawlingsSaunders1991, KinoKawakatu2005, GodfreyShabala2013, TurnerEA2018b, HardcastleCroston2020, TurnerShabala2023}. Our technical setup follows \citet{YatesJonesEA2021}: the jets are injected conically at a height of 0.5\,kpc and initial radius 0.1\,kpc, with a half-opening angle of 15 degrees; such jets are likely to retain their FR-II morphology following collimation \citep{Alexander2006, KrauseEA2012}. We refer the reader to \citet{YatesJonesEA2021} for technical details.

All our jets inflate radio galaxies several 100\,kpc in size at switch-off. We follow the remnant phase of evolution, and find (Section~\ref{sec:results}) that the synchrotron emission in this phase fades rapidly. The second suite of simulations therefore explores re-acceleration of cosmic ray electrons by passage of a plane parallel shock, creating a ``phoenix'' phase of radio emission. We explore four shock orientations: a normal shock (i.e. a plane parallel shock perpendicular to the jet axis), and three shocks inclined at 20, 45 and 70 degrees to the normal. The microphysics of DSA is complex, with shock strength, orientation, and the fraction of energy imparted to the electrons all potentially playing a role in determining particle acceleration efficiency \citep[e.g.][]{BoessEA2023}. We follow the approach of \citet{Kang2020} and assume a power-law energy distribution for remnant electrons revived by shocks, with the normalisation set by the total number of electrons injected during the active jet phase (see \ref{sec:AppendixRadioEmission}). As shown by \citet{Kang2020}, both a power law energy distribution and only a weak dependence of emissivity on the shock Mach number are found in semi-analytic DSA models for shocks with Mach number exceeding three, as in our simulations. In this aspect, our re-acceleration model is less dependent on unknown physics, and hence simpler, than the paradigm involving {\it in situ} shock acceleration of thermal electrons. In addition to a much weaker dependence on shock parameters \citep{Kang2020}, our fossil electron model also avoids the so-called pre-acceleration problem, in which low energy electrons cannot repeatedly cross the shock to undergo repeated acceleration due to their small gyroradii (see \citealt[e.g.][]{MalkovDrury2011,vanWeerenEA2016,KangEA2019} for details, and \citealt{CaprioliSpitkovski2014,RyuEA2019} for possible solutions).

\begin{table*}[ht]
\small
\centering
	\begin{tabular}{ccccccc} \hline \hline
	Simulation code & Type of & Single jet power & $t_{\rm on}$ & Domain & Grid cells & Resolution (kpc) \\
	& simulation & (W) & (Myr) & (kpc$^3$) & ($N_x \times N_y \times N_z$) & at 100, 1000 kpc \\
	\hline
 	Q38-t50 & active and remnant & $10^{38}$ & 50 & $1000^3$ & $290^3$ & 10, 10 \\ 
	\hline \hline
	Simulation code & Type of & Single jet power & $t_{\rm on}$ & Shock angle & Shock speed & $t_{\rm shock}$ \\
	& simulation & (W) & (Myr) & (degrees) & (km\,s$^{-1}$) & (Myr) \\
	\hline
	Q38-t50-s90 & normal shock & $10^{38}$ & 50 & $90$ & $3000$ & {280}  \\ 
	Q38-t50-s70 & 70 deg oblique shock & $10^{38}$ & 50 & $70$ & $3000$ & {280}  \\ 
	Q38-t50-s45 & 45 deg oblique shock & $10^{38}$ & 50 & $45$ & $3000$ & {280}  \\ 
	Q38-t50-s20 & 20 deg oblique shock & $10^{38}$ & 50 & $20$ & $3000$ & {280}  \\
	\hline
	\end{tabular}
\caption{Simulations. $t_{\rm shock}$ refers to the approximate age of the system at which the shock front reaches the remnant lobe.}
\label{tab:sims}
\end{table*}

Each active jet simulation required approximately 200k CPU hours on the University of Tasmania's {\it kunanyi} HPC cluster; remnant and shock simulations are significantly cheaper.

\section{Results}
\label{sec:results}

\begin{figure*}
\begin{center}

\includegraphics[]{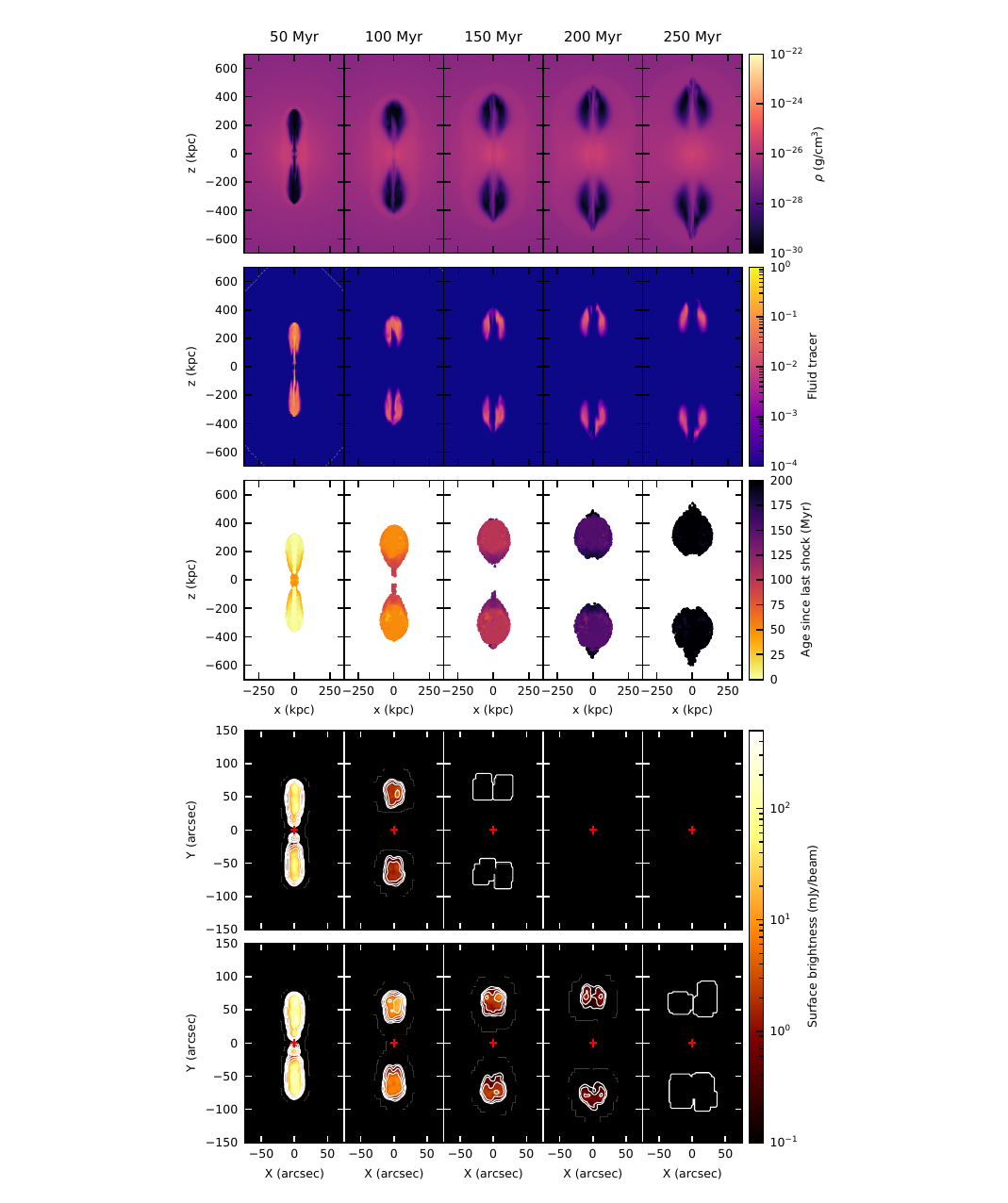}

\caption{Remnant lobes fade rapidly after the jet switches off. Hydrodynamic quantities and synthetic synchrotron emission in the plane of the sky for the Q38-t50 simulation at $z=0.3$. Columns show snapshots every 50 Myr; the left panels correspond to the switch-off time of 50 Myr; subsequent panels show remnant evolution. Top row: mid-plane density. Second row: mid-plane jet tracer. Third row: particle age since last shock; youngest particles are plotted on top. Fourth row: integrated surface brightness at 1 GHz, viewed in the plane of the sky and convolved with a 6 arcsecond beam FWHM. Contours are at 0.1, 0.3, 1, 3, and 10 mJy\,beam$^{-1}$. Bottom row: surface brightness at 200 MHz, convolved to the same beam. 1 arcsec = 4.5 kpc at the simulated redshift $z=0.3$, hence all plots are shown on the same spatial scales.}
\label{fig:sb_remnantFade}
\end{center}
\end{figure*}

Our choice of simulation parameters, specifically the low jet density, high speed and narrow jet opening angle, ensure that all jets considered in this work inflate lobes with characteristic edge-brightened FR-II morphology (see e.g. review by \citealt{TurnerShabala2023}). The narrowness of injected jets means that they have sufficient forward thrust to propel the terminal shock past the jet recollimation shock. Because the jets are light, the collimation is done by the cocoon inflated via backflow from the jet termination shock, rather than by the external medium; this produces narrow jets \citep{KomissarovFalle1998, Krause2005, Alexander2006, KrauseEA2012}, consistent with observations of FR-II radio galaxies. Figure~\ref{fig:sb_remnantFade} shows the evolution of moderate power ($10^{38}$\,W) relativistic ($v_{\rm j}=0.95c$) jets in the Q38-t50 simulation. Rows show relevant variables: density, jet tracer, jet particles, and synchrotron emissivity at 1000 and 200 MHz. At 50\,Myr (left column), a 550\,kpc radio galaxy with clear FR-II morphology is seen, including edge-brightened lobes and narrow jets. If ORCs are radio galaxies viewed close to end-on, they must be seen in the remnant phase to avoid emission from the jet head region, which would produce extended bright emission near the geometric centre of the ORCs; we note that this argument does not depend on the morphology (FR-I or FR-II) of the ORC progenitor.
\footnote{While edge-brightened lobes inflated by cosmic ray proton-dominated jets may appear as circles when viewed end-on \citep{LinYang2024}, for the majority of orientations such models predict edge-brightened structures which are inconsistent with observations of the radio galaxy population.}
Columns in Figure~\ref{fig:sb_remnantFade} show the dynamical and radiative evolution of the remnant lobes, starting at the jet switch-off time of 50\,Myr.

\subsection{Rapid fading of remnant lobes}
\label{sec:remnantDynamics}

A key tenet of this paper is that, in the absence of particle re-acceleration, the dynamical evolution of remnant radio lobes required to give the observed ORC morphology is much slower than the rate at which the lobe synchrotron emission fades. We now present numerical and analytical results in support of this argument.

As the remnant lobes evolve, the channel formerly occupied by the jet is excavated (top two rows of Figure~\ref{fig:sb_remnantFade}), and the lobes slowly transition to a toroidal morphology as predicted by \citet{ChurazovEA2001}. However, this transition is slower than both the decay of the strong bow shock ahead of the lobes, clearly seen in the density maps even at 150\,Myr; and the rapid fading in surface brightness (bottom two rows). These results are consistent with recent literature \citep{Hardcastle2018a, ShabalaEA2020}, and can be understood as follows.

\subsubsection{Fading timescale}
\label{sec:fadingTimescale}

The typical remnant fading time is $t_{\rm fade} \approx -\frac{\gamma}{d\gamma / dt}$, where $\gamma$ is the characteristic Lorentz factor emitting at observing frequency $\nu$. The radiative loss rate term in the denominator is $\frac{d\gamma}{dt}=-\frac{4}{3} \frac{\sigma_T}{m_e c} \gamma^2 (u_B + u_{\rm CMB})$, where $u_B$ is the lobe magnetic field energy density and $u_{\rm CMB} = 4.17 \times 10^{-14} (1+z)^4$\,J\,m$^{-3}$ is the energy density of CMB photons. This expression is a lower limit on the loss rate, since it ignores any adiabatic expansion which can be important to lobe evolution, particularly in the active phase \citep{KaiserEA1997, TurnerShabala2015, ShabalaGodfrey2013, Hardcastle2018a}. The fading timescale is therefore

\begin{equation}
t_{\rm fade} \leq \frac{3}{4} \frac{m_e c}{\sigma_T} \gamma^{-1} (u_B + u_{\rm CMB})^{-1}
\label{eqn:t_fade}
\end{equation}

At low redshift, the magnetic field energy density dominates Equation~\ref{eqn:t_fade}, with typical field strengths at several $\mu$G level \citep{HardcastleEA2016, InesonEA2017, TurnerEA2018a}. An estimate for the lobe field strength of $B \gtrsim \left( 2 \mu_0 \eta_{\rm eq} \eta_{\rm op} p_x \right)^{1/2} = 8.7 \times 10^{-10} (\eta_{\rm eq} \eta_{\rm op}) ^{1/2}$\,Tesla is obtained from the constraint that FR-II lobes do not suffer significant entrainment \citep{CrostonHardcastle2014, InesonEA2017}, and thus the lobe pressure should be approximately comparable to the thermal pressure in the external gas, $p_x \gtrsim 3 \times 10^{-13}$\,Pa in our simulated environments. The factor $\eta_{\rm eq} \equiv u_B / p$ \citep{HardcastleEA2016, InesonEA2017, TurnerEA2018b} is the ratio of magnetic field energy density to pressure in the lobes; and $\eta_{\rm op} \equiv p_l / p_x \approx 1-3$ \citep{CrostonEA2005, InesonEA2017, HardcastleCroston2020} is the overpressure factor of the lobes with respect to the surrounding gas. In our simulations we adopt $\eta_{\rm eq}=0.03$, corresponding to lobe magnetic fields of $\sim 10$\,$\mu$G at the time the powerful jets switch-off, and fading to a few $\mu$G level in the remnant phase, consistent with observations \citep{CrostonEA2018, KnuettelEA2019}. Hence the factor $ (\eta_{\rm eq} \eta_{\rm op}) ^{1/2}$ is of order unity. 

The minimum Lorentz factor of electrons required to produce emission at a frequency $\nu$ is $\gamma = \left( \frac{\nu}{\nu_L} \right)^{1/2}$ where $\nu_L = \frac{e B}{2\pi m_e}$ is the Larmor frequency (in SI units). Using typical scalings above, $\gamma = 6.4 \times 10^{3} \left( \frac{\nu}{1\,{\rm GHz}} \right)^{1/2} \left( \frac{u_B}{3 \times 10^{-13}\,{\rm Pa}} \right)^{-1/4}$ and the fading timescale is $t_{\rm fade} \leq 51\,{\rm Myr}  \left( \frac{\gamma}{6.4 \times 10^{3}} \right)^{-1} \left( \frac{u_B + u_{\rm CMB}}{3 \times10^{-13}\,{\rm Pa}} \right)^{-1}$.

\subsubsection{Dynamical timescale of buoyant bubbles}
\label{sec:buoyantTimescale}

The minimum timescale associated with the transformation of a remnant lobe to a torus (and hence an ORC when viewed close to head-on) is the time for cocoon expansion to slow down to subsonic velocities, plus the time for the morphological transformation. We now show that this timescale is significantly longer than the remnant fading timescale.

If ORC progenitors are large radio lobes, these must be inflated by moderate-to-high power jets for the following reasons. First, inflation of large (several hundreds of kpc) lobes in a reasonable (hundreds of Myr; \citealt{AlexanderLeahy1987, HarwoodEA2013, HardcastleEA2019}) time requires a substantial supersonic lobe expansion phase, which can only be provided by jets with high kinetic power \citep{BegelmanCioffi1989, KaiserAlexander1997, HardcastleKrause2013, TurnerShabala2023}; such a supersonic phase is also necessary for the lobes to not become substantially entrained by the external gas. Second, populations studies suggest that lower power sources are typically short-lived \citep{HardcastleEA2019}, and hences unlikely to produce very large sources.

The integrated radio luminosity of large sources declines with source size due to a combination of synchrotron, adiabatic and inverse Compton losses; in order to be visible (even when compressed into ring-like structures) these sources must therefore be powered by a large mass flux along the jet. This is an important consideration, because as pointed out by \citet{NoltingEA2019} the timescale for the transition to the buoyant rising bubble phase can be significant. We now show that, for parameters typical of powerful radio sources, this is substantially longer than the bubble fading time.

For source of size $D_s$ at switch-off time $t_s$, the remnant grows as \citep{KaiserCotter2002}
\begin{equation}
D(t) = D_s \left( \frac{t}{t_s} \right)^{2/(2-\beta+3\gamma_C)}
\label{eqn:KCdynamics_full}
\end{equation}
where the external medium density profile is $\rho_x(r) = \rho_0 \left( \frac{r}{r_0} \right)^{-\beta}$, and $\gamma_C = 4/3$ for a relativistic cocoon.
For our simulated cluster environment, $\beta \approx 1$ at radii well beyond $r_0 \approx 30$\,kpc, and hence $D(t) = D_s \left( \frac{t}{t_s} \right)^{2/5}$.

Hence the transition to subsonic expansion happens at time $t_b = t_s \left( \frac{\dot{D}_s}{c_s} \right)^{5/3}$. For the Q38-t50 simulation (Figure~\ref{fig:sb_remnantFade}), the observed cocoon size at switch-off $t_s=50$\,Myr is $D_s \approx 320$\,kpc, and the bow shock expansion speed at switch-off is $\dot{D}_s \approx 1.5 \times 10^6$\,km\,s$^{-1}$. This yields an expected $t_b \approx 120$\,Myr, consistent with the mildly supersonic forward velocities observed in the 100 Myr snapshot, and subsonic velocities in the 150 Myr snapshot in Figure~\ref{fig:sb_remnantFade}.

In the active phase, the relationship between single jet kinetic power $Q$, source age $t$, and size $D$ is given by \citep[e.g.][]{KaiserAlexander1997}
\begin{equation}
D(t) \propto \left( \frac{Q t^3}{\rho_0 r_0^5} \right)^{3/(5-\beta)} \approx \left( \frac{Q t^3}{\rho_0 r_0^5} \right)^{3/4}
\label{eqn:KAdynamics}
\end{equation}
For a given environment we therefore have $D(t) \propto (Q t^3)^{3/4}$ and $\dot{D} \propto Q^{3/4} t^{5/4}$ in the active phase.

Equations~\ref{eqn:KCdynamics_full} and \ref{eqn:KAdynamics} now predict the dependence of the buyoancy timescale on jet parameters, $t_b \propto Q^{5/4} t_s^{37/12}$. More powerful, longer-lived jets will take longer to reach the buoyant phase. For example, with reference to the Q38-t50 simulation, we expect a more powerful ($Q=10^{40}$\,W), shorter-lived ($t_{\rm on}=10$\,Myr) jet to enter the buoyant phase 2.2 times later, at around 260 Myr.

These timescales $t_b$ are all much longer than the fading timescale of the bubbles calculated in Section~\ref{sec:fadingTimescale}. Figure~\ref{fig:sb_remnantFade} confirms that remnants fade well before the development of toroidal structures characteristic of old remnants.

\subsection{Transient features in powerful backflows}
\label{sec:backflowORCs}

In principle, it may be possible for an ORC to form if the timescale $t_{\rm evac} \sim \frac{r_{\rm ORC}}{M_x c_s}$ associated with the evacuation of a cavity is shorter than the fading timescale. Here, $M_x$ is the Mach number of the flow with respect to the ambient medium, $c_s \sim 900$\,km\,s$^{-1}$ is the ambient sound speed, and $r_{\rm ORC} \approx 200$\,kpc is the characteristic radius of the ORC (e.g. ORC1, \citealt{NorrisEA2022}). External Mach numbers $M_x>4$, corresponding to flow velocities in excess of $0.01c$ are required for this dynamical timescale to be shorter than $t_{\rm fade}$ (Equation~\ref{eqn:t_fade}).

Detailed analysis of simulations in this paper, however, shows that while fast ($>15,000$\,km\,s$^{-1}$) backflow from strongly overpressured hotspots in recently switched-off sources can indeed temporarily evacuate central cavities, these structures are exceptionally transient and not necessarily ring-like in morphology; furthermore, the emission is very patchy, and fades rapidly over several Myr. These characteristics are at odds with observations of ORCs 1, 4, 5 and 6 which show narrow, smooth rings of emission \citep{NorrisEA2021_orcs, NorrisEA2022, KoribalskiEA2021}.

\section{Revived remnant lobes}
\label{sec:phoenices}

Discussion in Section~\ref{sec:remnantDynamics} shows that, for toroidal remnants to be visible, the synchrotron-emitting electrons must be re-accelerated. We consider this scenario next.

The large sizes (hundreds of kpc; Figure~\ref{fig:sb_remnantFade}) of remnant lobes inflated by typical ($Q \sim 10^{38}$\,W) FR-II jets are comparable to cluster virial radii, and these remnants will be subject to cluster ``weather''. Recently, \citet{RajpurohitEA2019, RajpurohitEA2021, DominguezFernandezEA2021} and \citet{WittorEA2021} have presented evidence for re-acceleration of fossil remnants by weak cluster shocks, for which there is ample observational evidence \citep{Planck2013, ChonEA2019}. In particular, \citet{DominguezFernandezEA2021} showed that the interaction of a cluster shock wave with a uniform Mach number with the ICM will naturally produce a range of Mach numbers, which in turn can produce radio spectra steeper ($\alpha \sim 1.1$) than expected in single-shock Diffusive Shock Acceleration models, and remarkably similar to ORC1 \citep{NorrisEA2022}. \citet{RussellEA2022} show these shocks are typically narrow and quasi-planar on scales of hundreds of kiloparsecs. In this section, we calculate the morphology of radio emission from remnant plasma reaccelerated by such weak shocks.

\subsection{Shock re-acceleration in a hydrodynamic simulation}

We initialise plane-parallel shocks in our simulations by setting the post-shock pressure and density as given by the hydrodynamic Rankine-Hugoniot conditions,

\begin{eqnarray}
p' = p \left( \frac{2 \Gamma_x M^2 - (\Gamma_x - 1)}{\Gamma_X + 1} \right) = \frac{5 M^2- 1}{4} p \nonumber\\
\rho' = \rho \left( \frac{(\Gamma_x+1)M^2}{(\Gamma_X-1) M^2 + 2} \right) = \frac{4 M^2}{M^2 + 3} \rho
\end{eqnarray}
Here, $p$ and $\rho$ refer to the unshocked ambient medium, and $P'$ and $\rho'$ to the shocked quantities. For a fiducial shock travelling at $3\,000$\,km\,s$^{-1}$ (Table~\ref{tab:sims}), corresponding to a Mach number of $M=3.3$ (see Section~\ref{sec:normalShocks}), we get a factor of 13.3 increase in pressure, and a factor of 3.1 increase in density, in the post-shock material.

These quantities, together with the relevant shock velocity, are implemented as a user-defined boundary condition in PLUTO at all post-shock locations. The initial location of the shock front is such that the shock first reaches the lobe at approximately 400\,Myr.

\subsection{Shocks normal to the jet axis}
\label{sec:normalShocks}

\begin{figure*}
\begin{center}

\includegraphics[]{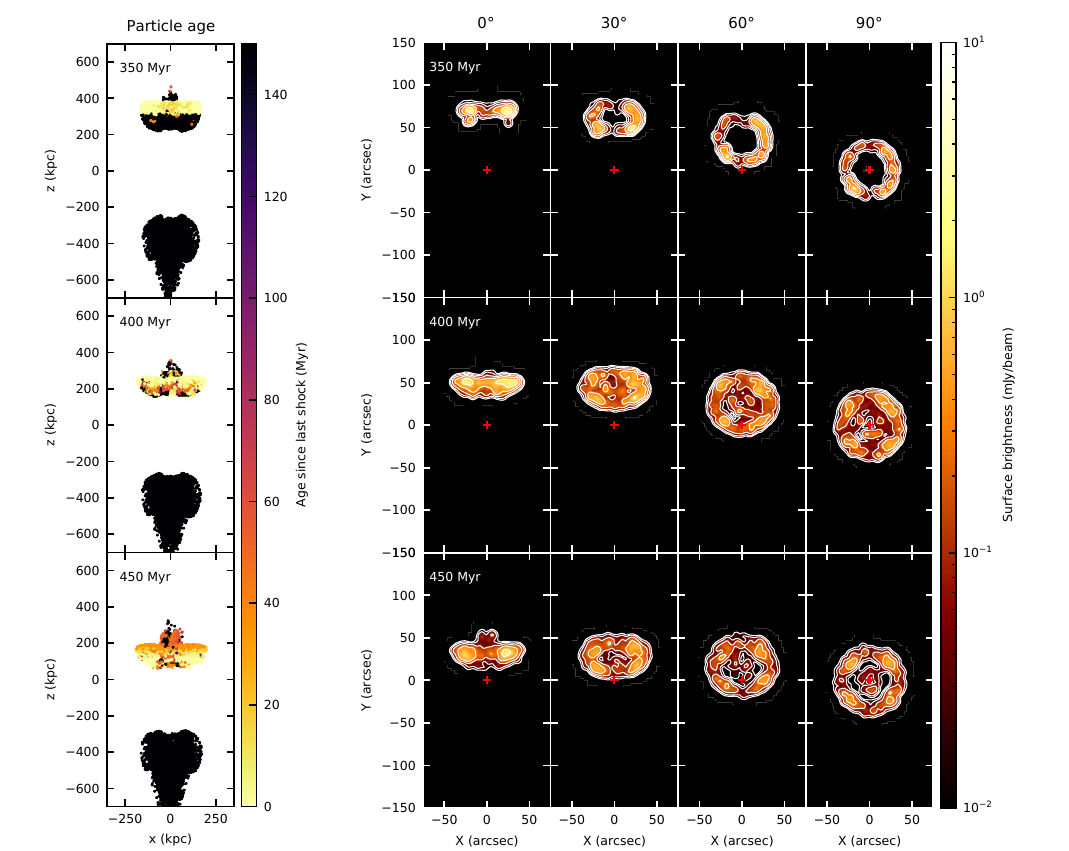}

\caption{Re-energized remnant radio lobes for different observing geometries. Lobes are inflated by a $10^{38}$\,W jet, active for 50\,Myr in a cluster environment; then evolve buoyantly until they are impacted by a plane-parallel normal shock, travelling at 3\,000 km\,s$^{-1}$ in the negative z-direction. Rows represent three different times since the onset of the shock. Columns from left to right are: (1) time since last shock for simulated lobe particles, as viewed in the plane of the sky; and projected radio surface brightness at 1.0 GHz at viewing angles of (2) 0 degrees (i.e. in the plane of the sky); (3) 30 degrees; (4) 60 degrees; and (4) 90 degrees (i.e. ``down the barrel'' of the switched off jet). Contours are at 0.01, 0.03, 0.1, 0.3, and 1 mJy\,beam$^{-1}$, and synthetic radio emission is convolved to a 6 arcsec FWHM beam. Circular, or quasi-circular rings of emission are clearly seen for angles inclined by 45 degrees or more to the line of sight. Ellipses are rare because of the fast (in terms of viewing angle) transition between ring and ``linear relic'' morphologies. The host galaxy is at the centre of the image in all cases.}
\label{fig:Q38-t50-shock-normal_timeSeries}
\end{center}
\end{figure*}

We first investigate the effects of a plane parallel shock oriented perpendicular to the jet axis; for jets propagating in the z-direction, our shock therefore lies in the $x$-$y$ plane. We adopt a shock speed of 3\,000\,km\,s$^{-1}$, corresponding to a Mach number number of 3.3 representative of cluster shocks (e.g. \citet{VazzaEA2016}, \citet{WittorEA2021}, and references therein). Figure~\ref{fig:Q38-t50-shock-normal_timeSeries} shows that such shocks are efficient at both compressing and ``lighting up'' particular sections of the remnant torus. For most observing orientations, the radio emission will appears either as a relic (for viewing angles $\lesssim 30^\circ$), or as a circular ring (for viewing angles $\gtrsim 45^\circ$). We provide a more detailed gallery of possible phoenix morphologies for both normal and non-normal shocks in \ref{sec:AppendixMoreAngles}.

\begin{figure}
\begin{center}

\includegraphics[]{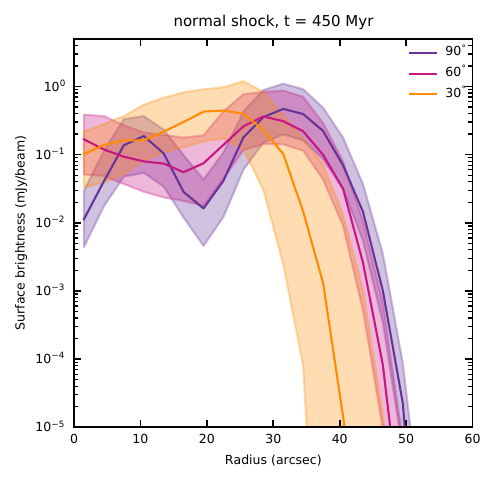}

\caption{Circularity of rings for different viewing angles, following passage of a normal shock. Lines show median surface brightness at 1 GHz, shaded region shows interquartile range for three viewing angles in Figure~\ref{fig:Q38-t50-shock-normal_timeSeries}. Quasi-circular rings are seen for 90 and 60 degree viewing angles. Departure from circularity (as given by the broadening of the interquartile range at a given radius) is observed for the 30 degree viewing angle.}
\label{fig:radialProfile_viewingAngle}
\end{center}
\end{figure}

Figure~\ref{fig:radialProfile_viewingAngle} shows the radial surface brightness distribution for different viewing angles. Even viewing geometries significantly offset from $90^\circ$ (i.e. not end-on) produce close to circular rings, as evidenced by a relative lack of broadening of the interquartile range at a given radius.

Importantly, for viewing angles other than $90^\circ$, the ORC host galaxy will not be at the centre of the ring. For example, shock-compressed lobes viewed at a $60^\circ$ angle will have the host (located at the origin in our simulation) {\it within} the ring. Some ORCs, such as ORC1, indeed have galaxies within the ring structure in addition to a galaxy at the centre of the ring \citep{NorrisEA2022}. However, the probability of the shock normal aligning perfectly with the jet axis is low, and hence below we consider phoenix morphologies for non-normal shocks.

We examine the circularity of phoenix emission and its relation to host galaxy location in more detail in Section~\ref{sec:singleORCs}.

\subsection{Switch-off time}

\begin{figure}
\begin{center}

\includegraphics[]{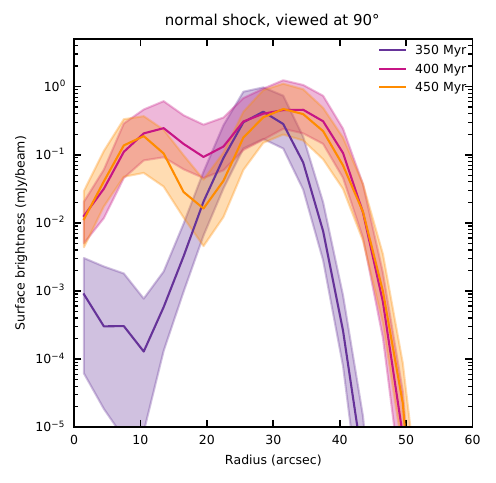}

\caption{Circularity of rings for different remnant ages. All snapshots are viewed head-on. Depending on how much lobe material has been swept up, both filled and edge-brightened rings can be produced.}
\label{fig:radialProfile_remnantAge}
\end{center}
\end{figure}

\begin{figure}
\begin{center}
\includegraphics[width=1.0\textwidth]{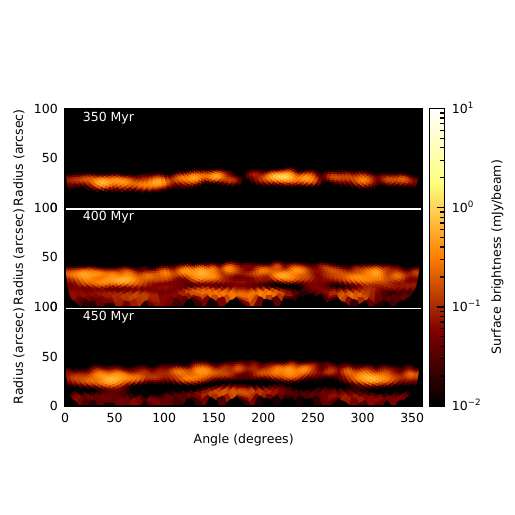}

\caption{Polar projection of radio phoenix emission at 1.0 GHz, at the same snapshots as in Figure~\ref{fig:radialProfile_remnantAge}. These are directly comparable to observations presented in Figure 4 of \citet{FilipovicEA2022}. }
\label{fig:radPlots}
\end{center}
\end{figure}

Figure~\ref{fig:Q38-t50-shock-normal_timeSeries} shows that ORC surface brightness and morphology, particularly the extent of diffuse emission, depends on the exact location of the shock. We quantify this effect in Figures~\ref{fig:radialProfile_remnantAge} and \ref{fig:radPlots}, which show that the observable features of the swept-up ring depend sensitively on shock location. At early times (350 Myr), the shock sweeps out a relatively narrow, edge-brightened ring. As the shock progresses, the synchrotron morphology evolves to a wider, brighter and centrally-filled ring (at 400 Myr), before fading to another edge-brightened ring at late times (450 Myr).

\subsection{Non-normal shocks}

We now examine the morphologies created by passage of a non-normal shock. 

\begin{figure*}
\begin{center}

\includegraphics[]{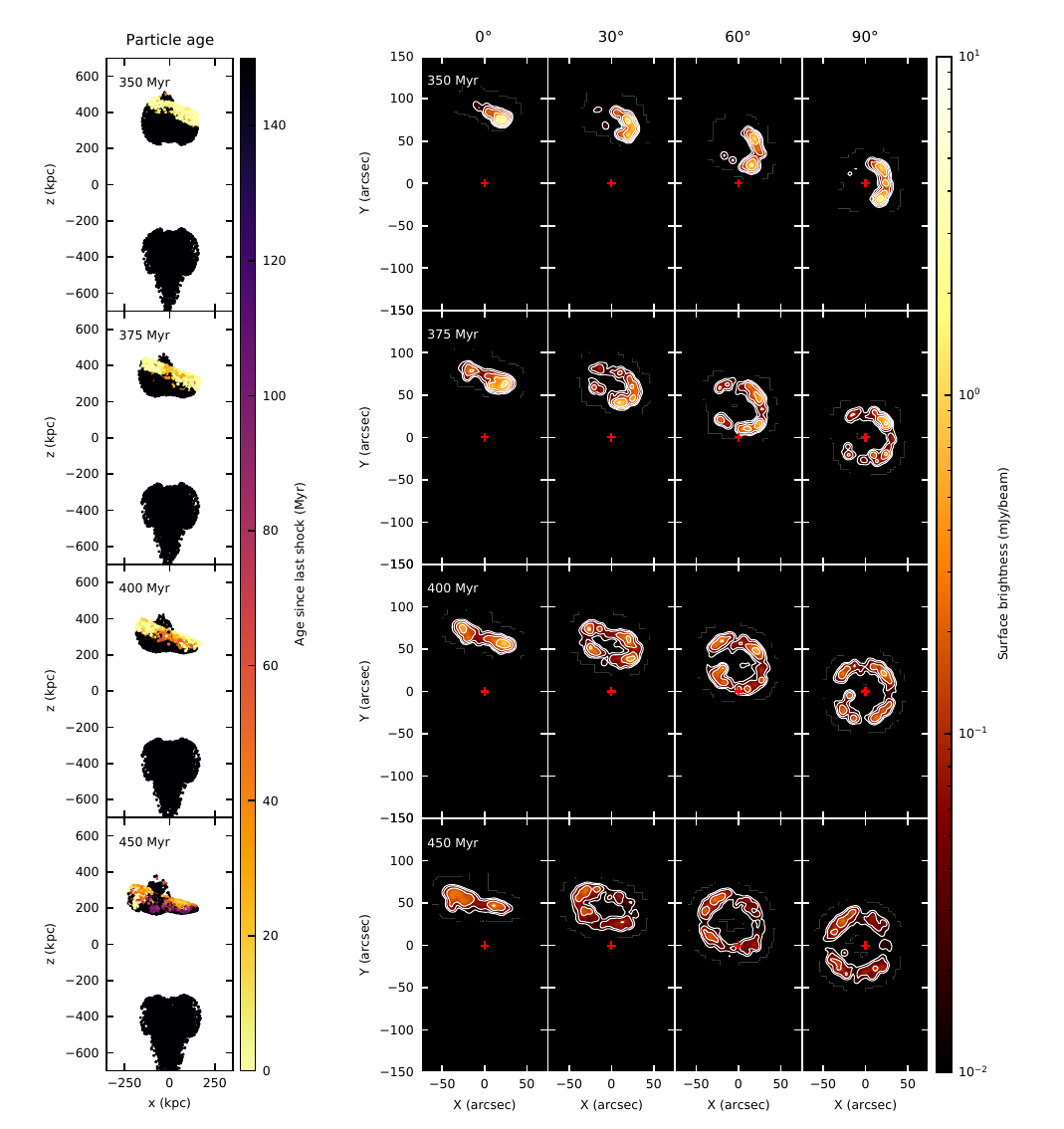}

\caption{As Figure~\ref{fig:Q38-t50-shock-normal_timeSeries}, but for a shock angled at 20 degrees to the normal. Arcs or rings will be seen depending on whether the shock has re-energized the full lobe cross-section.}
\label{fig:Q38-t50-shock-20deg_timeSeries}
\end{center}
\end{figure*}

\begin{figure*}
\begin{center}

\includegraphics[]{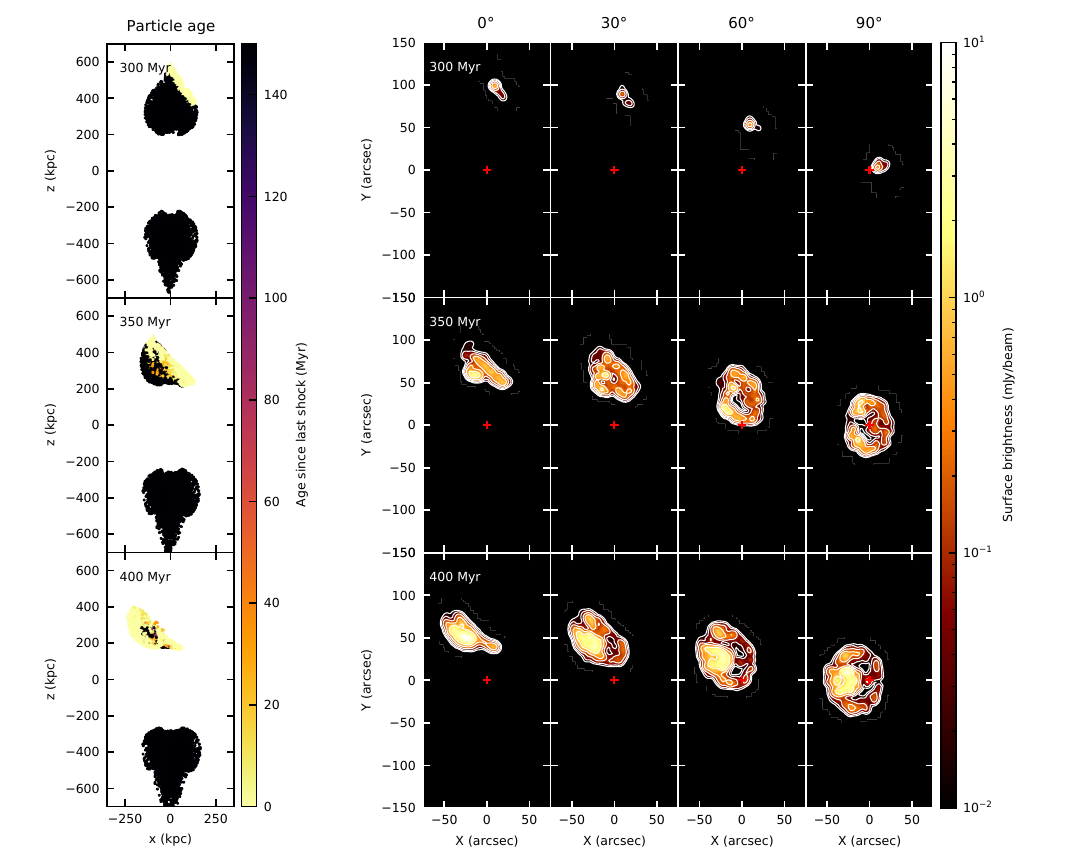}

\caption{As Figure~\ref{fig:Q38-t50-shock-normal_timeSeries}, but for a shock angled at 45 degrees to the normal.}
\label{fig:Q38-t50-shock-45deg_timeSeries}
\end{center}
\end{figure*}

\begin{figure*}
\begin{center}
\includegraphics[]{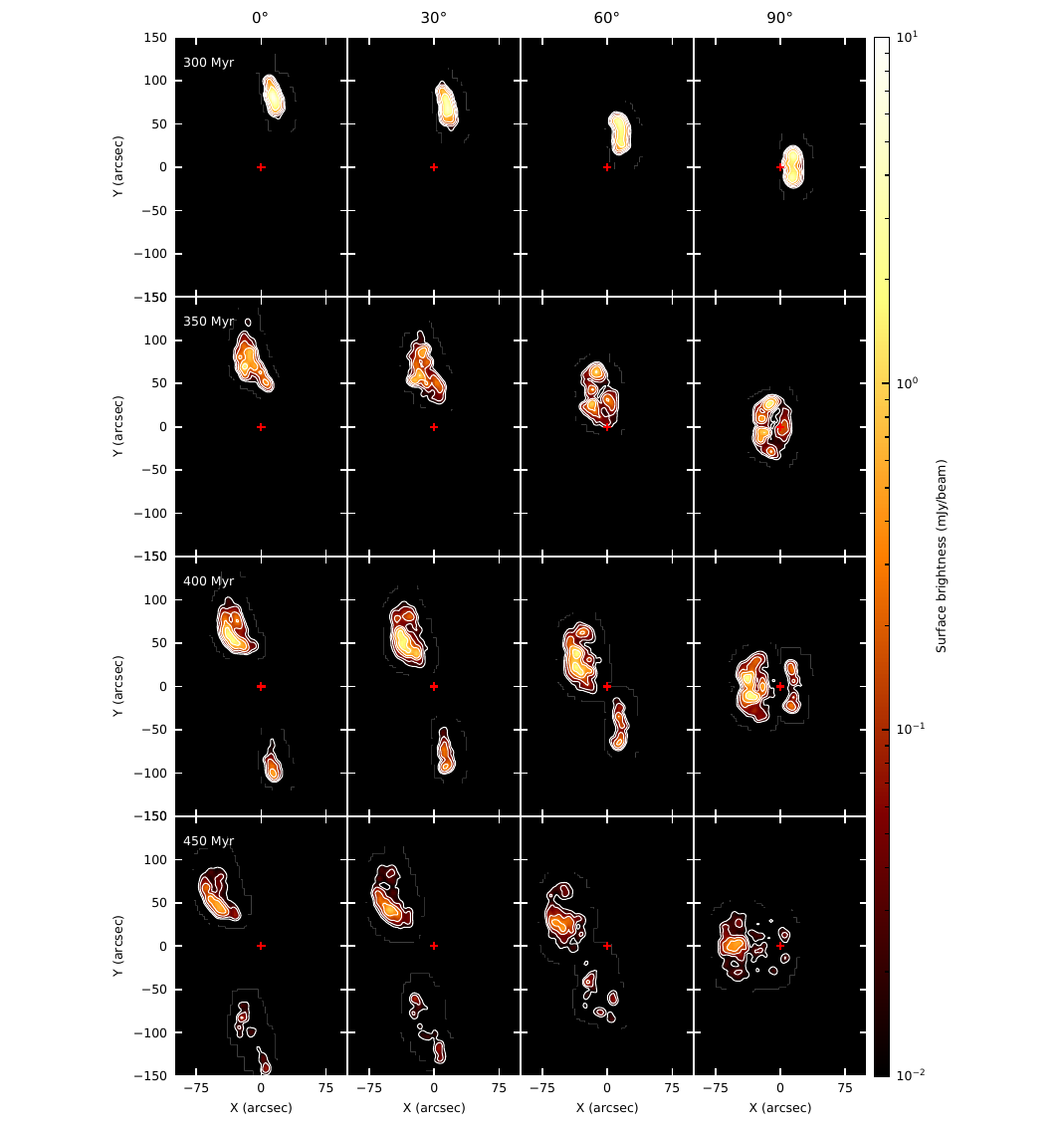}

\caption{As Figure~\ref{fig:Q38-t50-shock-normal_timeSeries}, but for a shock angled at 70 degrees to the normal, i.e. a quasi-parallel shock. No clear rings or ellipses are seen.}
\label{fig:Q38-t50-shock-70deg_timeSeries}
\end{center}
\end{figure*}

\begin{figure}
\begin{center}
\includegraphics[]{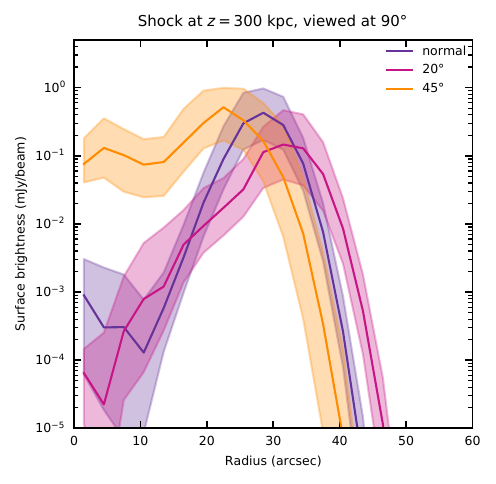}

\caption{Circularity of rings for different shock orientations. Snapshots are selected so that the shock mid-point is at coordinate $z \sim 300$\,kpc; this corresponds to the 350 Myr snapshot for the normal shock, 400 Myr snapshot for the quasi-normal 20 degree shock; and 350 Myr snapshot for the 45 degree shock. Quasi-circular rings can be produced even by non-normal shocks.}
\label{fig:radialProfile_shockAngle}
\end{center}
\end{figure}

Figure~\ref{fig:Q38-t50-shock-20deg_timeSeries} shows the evolution of a plane parallel shock inclined at 20 degrees to the normal, at several viewing angles; Figure~\ref{fig:Q38-t50-shock-45deg_timeSeries} shows a shock at 45 degrees to the normal; and Figure~\ref{fig:Q38-t50-shock-70deg_timeSeries} a shock at 70 degrees to the normal (i.e. a quasi-parallel shock). A comparison of characteristic radio phoenix morphologies is shown in Figure~\ref{fig:radialProfile_shockAngle}.

More complex shock geometries produce a qualitatively similar, but more nuanced picture: toroidal radio structures (seen as circular emission when viewed close to edge-on) still form at late times for shocks closer to normal than parallel (Figure~\ref{fig:Q38-t50-shock-20deg_timeSeries} and Figure~\ref{fig:Q38-t50-shock-45deg_timeSeries}), when the shock has interacted with the full lobe cross section. At earlier times, however, arc-like structures ``light up'' only the side of the lobe which has interacted with the shock; these arcs undergo significant evolution, in both morphology and surface brightness, as the shock progresses. Even for end-on observing geometries, non-normal shocks also produce less symmetric structures; this asymmetry increases at large viewing angles, and for larger shock angles. For the largest shock angles, such as the quasi-parallel shocks in Figure~\ref{fig:Q38-t50-shock-70deg_timeSeries}, no observable ellipses or circles are produced as circular symmetry is destroyed by the shock passage.

\section{Discussion}
\label{sec:discussion}

\subsection{Implications of the model}

The results presented in Section~\ref{sec:phoenices} show that remnant radio lobes revived by a shock passage may provide a plausible explanation for the ORC phenomenon. Diffuse and edge-brightened rings are seen for a wide range of shock geometries, viewing angles and shock ages.

A diversity of ORC geometries is possible, with younger shocks creating smaller, brighter, more filled (i.e. less edge-brightened) quasi-circular structures. For a given shock angle, elliptical ORCs will be more rare than expected from simple projection arguments, because of the relatively rapid transition -- with viewing angle -- from mushroom cap to quasi-circular geometry. A testable prediction of our model is that, for non-normal shocks, viewing angles sufficiently away from the line-of-sight should produce both more elliptical and more asymmetric (i.e. one side of the ring wider than the other) ORCs. Depending on surface brightness sensitivity, such structures may be seen as one-sided arcs (e.g. Figure~\ref{fig:Q38-t50-shock-20deg_timeSeries}).

Because our simulations do not include magnetic fields, we cannot make robust predictions for spatially resolved spectra in simulated ORCs. To first order, when the shock first encounters the lobe the newly-shocked relic electrons attain a (Mach number-dependent) spectral index reflecting their energy distribution; this spectrum gradually steepens as the electrons age on timescales comparable to the shock crossing time (cf. Equation~\ref{eqn:t_fade}). In principle, non-normal shock and viewing geometries may produce spectral gradients across the phoenix structure; details will depend on the detailed interaction between shock and remnant dynamics, and the structure of the lobe magnetic field which may become increasingly complex in the transition to the phoenix phase.

\subsection{Single ORCs}
\label{sec:singleORCs}

Rings of radio emission can be produced by quasi-normal shocks for a relatively broad range of viewing angles close to the line of sight. Because the jet axis and shock orientation are expected to be independent of each other, shock normal orientations exactly aligned with the jet axis such as in Figure~\ref{fig:Q38-t50-shock-normal_timeSeries} are unlikely. The probability of jet-shock angle orientation in the range $(\theta, \theta + d \theta)$ is proportional to $(1-\cos \theta) \, d\theta$, hence shocks aligned to within $10^{\circ}$ of the jet axis (as in Figure~\ref{fig:Q38-t50-shock-normal_timeSeries}) are 7.8 times less likely than shocks misaligned by $20^{\circ}$ (as in Figure~\ref{fig:Q38-t50-shock-20deg_timeSeries}), and 16 times less likely than shocks misaligned by $45^{\circ}$ (as in Figure~\ref{fig:Q38-t50-shock-45deg_timeSeries}). Quasi-parallel shocks (e.g. Figure~\ref{fig:Q38-t50-shock-70deg_timeSeries}) are more likely still, but these do not produce circular or elliptical post-shock emission. Therefore, the shocks most likely to produce ORC-like structures are quasi-normal ones such as those in Figure~\ref{fig:Q38-t50-shock-20deg_timeSeries}.

This has important implications for the likely location of the ORC host galaxies. For normal shocks, hosts can be significantly offset from the ORC geometric centre while still retaining circular structure when viewed away from the line-of-sight; for example, in the $60^{\circ}$ viewing angle in Figure~\ref{fig:radialProfile_viewingAngle} the ORC host galaxy would be located inside the radio ring.

\begin{figure*}
\begin{center}
\includegraphics[]{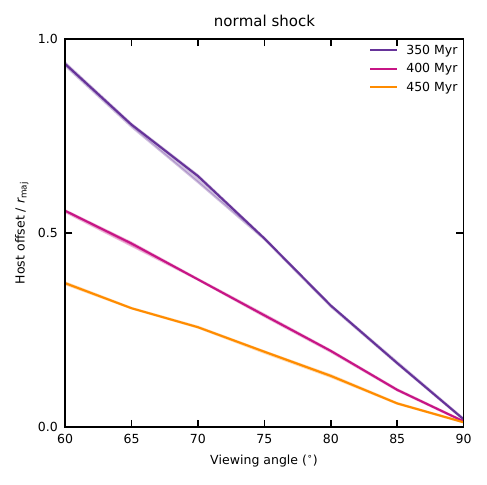}
\includegraphics[]{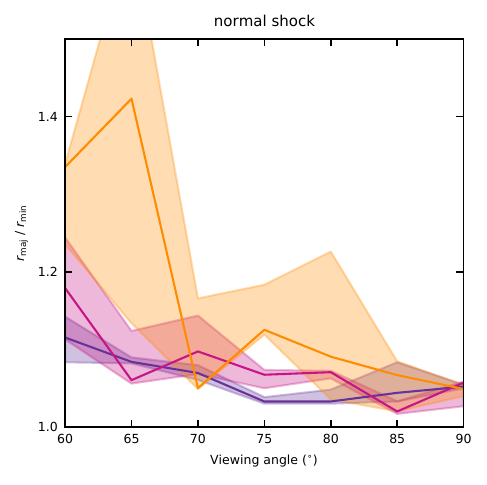}
\includegraphics[]{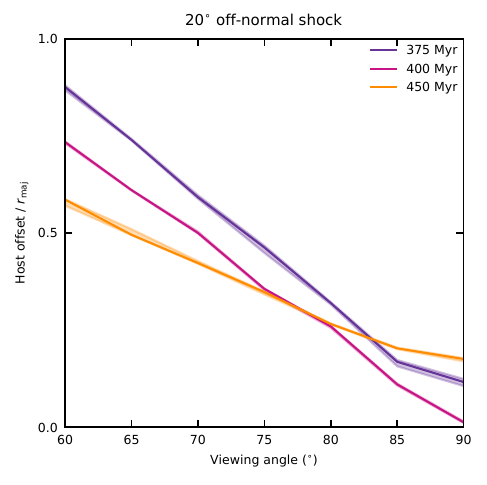}
\includegraphics[]{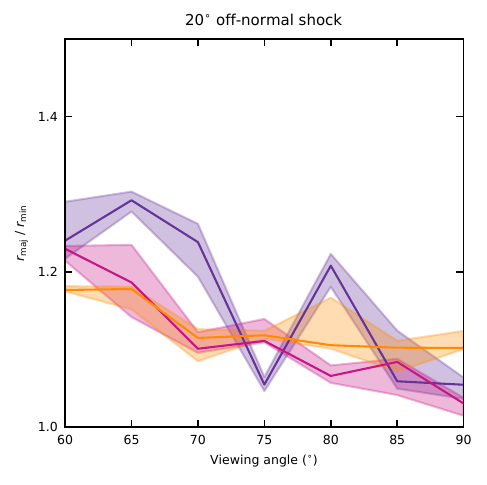}
\caption{Host offset from the geometric centre of emission (left panel) and circularity (defined as the ratio of semi-major to semi-minor axes of the synthetic radio emission, right panel) observed at a range of viewing angles. Solid lines denote medians for a range of simulated sensitivities (0.5-2 $\mu$Jy/beam for a 6 arcsec beam), and shaded regions show the interquartile range. Top row: normal shock. Bottom row: quasi-normal (20 degree offset) shock. Offset of the host galaxy from the ORC geometric centre increases with viewing angle away from the line of sight, while circularity decreases. For the quasi-normal shock, only observing geometries close to line of sight can produce highly circular structures.}
\label{fig:ellip_90deg}
\end{center}
\end{figure*}

We quantify this effect in Figure~\ref{fig:ellip_90deg}. For each normal and quasi-normal shock snapshot presented in Figures~\ref{fig:Q38-t50-shock-normal_timeSeries} and \ref{fig:Q38-t50-shock-20deg_timeSeries}, we calculate the centre of emission, and semi-major and semi-minor axes lengths $r_{\rm maj}$ and $r_{\rm min}$. The resultant offset of the centre of the radio emission from the host galaxy (located at the origin in our simulations), and the ratio $r_{\rm maj} / r_{\rm min}$, are plotted in Figure~\ref{fig:ellip_90deg}. While there is significant variability with shock angle and time of observation, some clear trends emerge. The host galaxy can be significantly offset from the centre of the radio emission, however this requires viewing angles some way away from the line of sight. For non-normal shocks -- which are more common -- such viewing angles also correspond to more elliptical structures, i.e. higher $r_{\rm maj} / r_{\rm min}$ ratios at viewing angles further away from $90^{\circ}$ in Figure~\ref{fig:ellip_90deg}. A requirement for the radio emission to be close to circular requires viewing angles close to the line of sight, and hence more central host galaxies.

\begin{figure*}
\begin{center}
\includegraphics[]{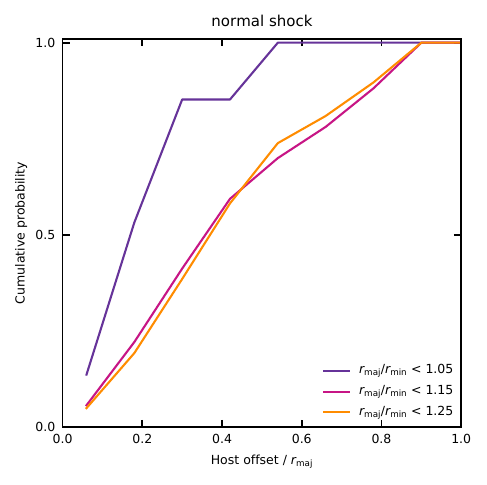}
\includegraphics[]{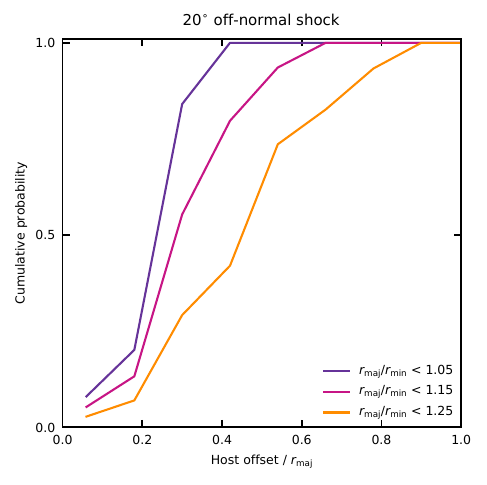}
\caption{Cumulative probability of the location of the host galaxy for a range of ellipticities. Left panel: normal shock; right panel: quasi-normal shock. The most circular structures ($r_{\rm maj} / r{\rm min} \approx 1$) will have host galaxies close to the geometric centre of the ORC.}
\label{fig:offset_prob}
\end{center}
\end{figure*}

In Figure~\ref{fig:offset_prob} we calculate the probability distribution of host galaxy locations for a range of $r_{\rm maj} / r_{\rm min}$ values, integrated over all viewing angles. For the most circular ORCs, the majority of host galaxies will be near the centre ($\lesssim 0.3 r_{\rm maj}$) of the radio emission; this trend is stronger for non-normal shocks (right panel), which are more common. 

These results provide a potential explanation for observations of ORCs 1, 4 and 5, where host galaxies are associated with the ORC geometric centre at high confidence \citep{NorrisEA2022}. The observed ORCs all have candidate host galaxies within 10 percent of the ring radius, which is more central than the predictions in Figure~\ref{fig:offset_prob}. We note that DSA will be more effective at accelerating the cosmic ray electrons in quasi-normal shocks \citep{BoessEA2023}, due to the expected topology of the magnetic field swept out by the lobes during the active phase \citep{HardcastleKrause2014,NoltingEA2019}; this will further enhance the observability of quasi-normal shocks. A more detailed exploration of simulated ORC ellipticities for a range of shock and viewing geometries, and larger samples of observed ORCs with host galaxies, are required to examine this further. A testable prediction of our model is that host location will move away from the ORC geometric centre with increased ellipticity. Less circular and edge-brightened ORC candidates \citep[e.g. ][]{GuptaEA2022} may be off-centre radio phoenixes.

\subsection{ORCs with lobes}
\label{sec:lobeORCs}

ORCs associated with radio lobes (e.g. ORCs 2 and 3, \citealt{NorrisEA2021_orcs}) are more challenging to explain in this model. A shock interacting with a lobe still emitting detectable synchrotron radiation will only produce strongly edge-brightened emission when viewed in projection after it has traversed the whole lobe -- before this time, the parts of the lobe closer to the host galaxy will still yield detectable radio emission near the ORC geometric centre. Because the lobes fade rapidly (Section~\ref{sec:fadingTimescale}), this scenario requires a fast shock reaching the remnant shortly after the jet switches off, then traversing it rapidly while the counterlobe is still visible, as proposed by \citet{NoltingEA2019}.

Figures~\ref{fig:Q38-t50-shock-20deg_timeSeries}, \ref{fig:Q38-t50-shock-45deg_timeSeries} and \ref{fig:Q38-t50-shock-70deg_timeSeries} suggest another plausible scenario. If the two remnant lobes are not anti-parallel, a single shock passing through both lobes will produce structures which look different in projection. An ORC plus lobe morphology will be seen at certain viewing angles if the axis of symmetry for one of the lobes but not the other is close to the shock normal. Wide Angle Tailed radio sources (WATs) are common in clusters \citep{ODeaBaum2023}, and Mpc-scale cluster shocks \citep[e.g.][]{RussellEA2022} can shock both lobes; hence this scenario is plausible. We note that in both ORCs known to be associated with a radio lobe \citep{NorrisEA2021_orcs,MacgregorEA2023}, the two sets of radio structures are not co-linear, consistent with our hypothesis. Alternative mechanisms such as precessing jets \citep{NoltingEA2023} are also possible: jet precession has been reported in powerful radio galaxies \citep{KrauseEA2019}, and ring-like structures attributed to jet precession have been observed in 3C\,310 \citep{KraftEA2012} and NGC\,6109 \citep{RawesEA2018} albeit on much smaller scales.

\subsection{Detectability of ORCs}
\label{sec:fracORCs}

Source age, shock orientation and viewing geometry all play a role in setting the observed morphology of the resulting radio phoenix. While detailed analysis of simulated source morphology is beyond the scope of this paper, we now provide estimates of the expected ORC fractions.

For quasi-normal shocks such as those shown in Figures~\ref{fig:Q38-t50-shock-normal_timeSeries_moreAngles} and \ref{fig:Q38-t50-shock-20deg_timeSeries_moreAngles}, circular structures appear at viewing angles $\geq 60^\circ$, ellipses for viewing angles $30^\circ - 60^\circ$, and unstructured morphologies for viewing angles closer to side-on than approximately $30^\circ$. The probability of each range in viewing angle is proportional to the subtended solid angle $\Delta \Omega = 2\pi \Delta \left( \sin^2 \theta \right)$. The surface brightness is similar for the different viewing angles. Hence approximately 13 percent of remnants revived in quasi-parallel shocks are expected to be seen as circles, 37 percent as ellipses, and half as neither.

The fraction of ORCs among all shock-revived remnants will be much lower. Circular and elliptical structures are not visible for quasi-parallel shocks, such as shown in Figure~\ref{fig:Q38-t50-shock-70deg_timeSeries}. For intermediate shock angles (Figure~\ref{fig:Q38-t50-shock-45deg_timeSeries}), even when ellipsoidal structures are produced there can be significant brightness asymmetry between different sides of the radio phoenix. Taking $30^\circ$ as the largest off-normal angle that can produce detectable ellipsoidal structures, only 3 percent of radio phoenixes are expected to be either ORCs or ellipses. The real fraction is likely to be even lower due to asymmetries in realistic remnant environments and cluster weather.

\subsection{Expected ORC densities}
\label{sec:numberCounts}

We can test the plausibility of the ORCs as radio galaxy phoenixes hypothesis, by estimating the expected sky density of these objects.

The expected number of ORCs in a solid angle $\Omega$ on the sky is given by

\begin{equation}
N_{\rm ORC} = \int_0^{z_{\rm max}} dz\, \Omega D_L(z)^2 \left( \frac{d D_L(z)}{dz} \right) \Phi_{\rm RG} f_{\rm shock} f_{\rm orient} f_{\rm life}
\label{eqn:N_orcs}
\end{equation}

where $D_L$ is the luminosity distance, $\Phi_{\rm RG}$ is the volume density of ORC progenitors, assumed to be powerful radio galaxies; $f_{\rm shock}$ is the fraction of radio galaxy remnants revived by a shock; $f_{\rm orient}$ is the fraction of sources with favourable orientation; and $f_{\rm life} = t_{\rm visible} / t_{\rm active}$ is the ratio between the time for which the remnant is visible and the age of the jet which produced the remnant. The limit of the integral, $z_{\rm max}$, corresponds to the maximum redshift to which a typical ORC can be observed, and depends on ORC morphology (including shell thickness) as well as survey surface brightness limits.

We can now make a (very approximate) estimate for the expected number of ORC1-like sources in the EMU Pilot Survey. Below, we adopt the following survey parameters (see \citealt{NorrisEA2021_emu}): surface brightness sensitivity at 1\,GHz of  $\Sigma_{\rm min}=90$\,$\mu$Jy/beam (corresponding to approximately $3\sigma$), $\Omega=0.26$\,sr (corresponding to 270 sq. deg.), and beam Full Width at Half Maximum (FWHM) of 12 degrees.

Relevant parameters for the best-studied Odd Radio Circle,  ORC1, are: ASKAP surface brightness $\Sigma_{\rm ORC}=0.13$\,mJy/beam, $z_{\rm ORC}=0.55$ \citep{NorrisEA2021_orcs}, and ring width of 40 kpc (corresponding to ORC1 being marginally resolved with MeerKAT; \citealt{NorrisEA2022}). For the assumed sensitivity, the maximum observable redshift of ORC1 is $z_{\rm max} \sim 0.6$.

For representative scaling parameters we adopt $\Phi = 10^{-7}$\,Mpc$^{-3}$, representative of the number density of powerful radio galaxies in the local Universe \citep{HardcastleEA2019}, a substantial fraction of which are associated with clusters \citep{CrostonEA2019}. From results in this work, we estimate $f_{\rm orient} \sim 0.01$, i.e. 1 percent of visible remnants produce ring-like structures. The largest uncertainty comes from the term $f_{\rm life} f_{\rm shock}$. We adopt $0.05 \leq f_{\rm shock} \leq 0.65$, where the upper limit is set by the observed cluster merging rate \citep{CassanoEA2016} and the lower limit by the fraction of prominent bow shocks in clusters in cosmological simulations \citep{Lokas2023}; we note that the association between mergers and powerful radio galaxies \citep{RamosAlmeidaEA2012,KrauseEA2019} may increase $f_{\rm shock}$. Figure~\ref{fig:Q38-t50-shock-70deg_timeSeries} shows that shocked remnants are visible after 150 Myr for an initial jet active phase of 50 Myr, setting $f_{\rm life} \geq 3$. We therefore estimate $f_{\rm life} f_{\rm shock} \sim 1$, noting the large uncertainties on this parameter.

An order-of-magnitude estimate for the expected number of ORCs in the EMU Pilot Survey field is then $N_{\rm ORC} = 1.5 \left( \frac{\Phi_{\rm RG}}{10^{-7}\,{\rm Mpc^{-3}}} \right) f_{\rm shock} f_{\rm life} \left( \frac{f_{\rm orient}}{0.01} \right)$. Although these numbers appear plausible, and are broadly consistent with the handful of ORCs discovered so far by the EMU survey, they should of course be interpreted as order-of-magnitude estimates only due to large uncertainties in several parameters.

\subsection{Future outlook}

While capable of accurately following the dynamics of the full radio jet lifecycle, and making robust global predictions for synchrotron radio emission, our hydrodynamic simulations do not capture the complex processes responsible for turbulent re-acceleration of remnant plasma. Instead, we adopt a simplified approach of a single shock threshold (\ref{sec:PRAiSE}), and assign a single power-law in electron energy to particles assumed to be accelerated at such shocks. More sophisticated treatments of particle re-acceleration, including full MHD and capturing of the turbulent cascade, have recently been presented by \citet{DominguezFernandezEA2021} and \citet{WittorEA2019, WittorEA2021}. Their findings validate our approach: \citet{DominguezFernandezEA2021} show that magnetic fields are not significantly compressed by weak (Mach 2, i.e. similar to those adopted in this paper) shocks; while \citet{WittorEA2019} showed that re-acceleration of remnant plasma by sustained turbulence can lead to uniform spectral indices similar to those observed in ORC1 \citep{NorrisEA2022}. We note that, to re-accelerate a \emph{non-thermal} pool of electrons, as assumed in our model, shocks weaker than the canonical threshold Mach number of $\sqrt{5}$ are sufficient \citep{VinkYamazaki2014}.

Our approach of calculating radio emission in post-processing necessitates some assumptions. As discussed in Section~\ref{sec:radioEmission}, the most important is the low-energy cutoff for radiating particles, which is directly proportional to the normalisation of the radio emission (cf. Equation~\ref{eqn:Nrad}). Our purely hydrodynamic simulations neglect the role of magnetic fields; while not dynamically important in the momentum-dominated phase of lobe evolution \citep[e.g.][]{HardcastleKrause2014,EnglishEA2019}, magnetic field structure will affect the spatial distribution and and spectra of the radio emission. The sweeping up of magnetic field lines can also provide additional stability (along with gas viscosity) to remnant bubbles \citep{KaiserEA2005,ShabalaAlexander2009b}. On the other hand, our simulations may overestimate bubble stability by not having sufficient resolution to capture instabilities at the bubble surface, and not including complex gas and shock dynamics \citep[e.g.][]{DolagEA2023} in our idealised environments.

Our idealised assumption of plane-parallel shock geometry would be improved by instead considering radio lobe evolution in dynamical, cosmological MHD simulations \cite[e.g.][]{KangRyu2015, VazzaEA2016, VazzaEA2019, VazzaEA2021, HodgsonEA2021,DolagEA2023}. A crucial point to note, however, is that modelling the revived remnant dynamics requires an accurate description of the full jet lifecycle, because the dynamics of the remnant bubbles are sensitive to parameters describing the earlier, active phase of the jet-inflated lobes (see Section~\ref{sec:buoyantTimescale}). The current generation of cosmological simulations do not model relativistic jets as observed in powerful FR-II radio sources -- the putative ORC progenitors. New, hybrid codes involving both relativistic jets and cosmological initial conditions (e.g. {\texttt CosmoDRAGoN}, \citealt{YatesJonesEA2023}) will be required for this work.

Shock interactions with the remnant plasma may produce detectable X-ray emission. \citet{DolagEA2023} considered this issue in some detail for their shocked CGM model, and  concluded that this is plausible, but large uncertainties in environment properties make accurate predictions challenging. We defer a detailed investigation of this to future work.

Our simulations predict that only 1 percent of shock-lobe encounters will result in circular radio emission, and 2 percent in elliptical structures.  Because of the interest in ORCs, considerable efforts have been made to find new ones. However, the elliptical and other structures produced in these encounters also have a great deal to teach us about radio galaxy physics and the presence of shocks in the environment. Observational samples of a broad range of structures, some of which are illustrated in this paper, are required to explore these aspects. Compiling such samples may be challenging  because the host galaxies can be significantly offset from the radio emission; nevertheless they are likely to be worth the effort.

\section{Conclusions}
\label{sec:summary}

We have run hydrodynamic simulations of powerful, relativistic jets in cluster environments, and calculated synchrotron radio emission in post-processing, to test whether quasi-circular structures resembling Odd Radio Circles (ORCs) can be produced by radio lobes. Our main findings are as follows.

\begin{itemize}
\item Large ($>100$)\,kpc remnant lobes fade below the detection limit faster than their morphology evolves into a toroidal shape. Hence old remnants cannot form ORCs.
\item Strong backflow from powerful jets can, for a short time, evacuate a cavity near the jet head. Such structures, however, are too transient and diffuse to be consistent with the observed ORCs.
\item Passage of a moderate Mach number shock sweeps up and re-accelerates remnant lobe plasma, creating ORC-like structures for a range of shock and observing geometries. 
\item The majority of shock -- lobe interactions will not form either circular or elliptical structures.
\item Circular structures can be produced by the passage of a quasi-normal shock (angle between shock normal and the jet axis $\leq 20^\circ$), and viewing angle within approximately $30^\circ$ of the jet axis. 
\item The width, surface brightness and ellipticity of the shock-induced radio phoenix depends on shock age and angle, and observing orientation. Asymmetric structures, including arcs, are predicted for off-axis shocks, and shocks which have only interacted with part of a lobe. 
\item Circular ORCs with a geometric centre which is significantly offset from the host galaxy location require the shock normal to be closely aligned with the jet axis, and hence are rare. The more likely off-axis shocks produce circular structures with the host galaxy near the geometric centre, consistent with observations of known ORCs. Ellipsoidal radio structures are predicted to have host galaxies which are offset from the centre of the ellipse.
\item ORCs with lobes, such as ORCs 2 and 3 \citep{NorrisEA2021_orcs}, can be produced by shock passage through the lobes of a Wide Angle Tailed radio source.
\end{itemize}

In our radio phoenix model, ORCs can only be produced if the remnant radio lobes remain intact before the shock passage revives the non-thermal emission. \citet{BotteonEA2023} recently showed that radio lobes will be shredded by gas sloshing in clusters on a timescale of approximately 500\,Myr, corresponding to several sound-crossing times. This timescale is longer than the duration of the remnant phase in all our simulations -- which focus on powerful radio sources -- and hence our proposed mechanism is plausible; however this requirement becomes increasingly challenging for lower power jets in dynamic environments.

The radio phoenix model combines two well-established features of the cosmological structure formation paradigm: large-scale shocks and radio galaxy lobes. Our simulations show that ORC-like structures can be produced when these shocks interact with remnant lobes. \citet{DolagEA2023} recently showed that direct shock acceleration of thermal electrons on halo outskirts may also be able to produce similar ring-like structures. Both hypotheses are plausible; the prediction of a relationship between host galaxy offset and ORC ellipticity in our phoenix model may help distinguish between these two scenarios. To definitively answer whether or not a large number of ORCs are indeed radio galaxy phoenixes will require large, detailed samples of observed ORCs, together with ever more sophisticated magnetohydrodynamic simulations of remnant jets in realistic, dynamic environments. 

\section{Acknowledgements}

We gratefully thank the Tasmanian Partnership for Advanced Computing, funded and hosted by the University of Tasmania, for high performance computing resources. We are grateful to the anonymous referee for their insightful and constructive comments, which have improved the paper.


\bibliography{orc-shocked-remnant}

\appendix

\section{Simulated radio emission}
\label{sec:AppendixMoreAngles}

Radio phoenix geometries are shown below for a normal shock, quasi-normal shock at $20^\circ$, and a shock at $45^\circ$ to the normal, for several snapshots. These plots complement those in the main body of the paper by showing a larger number of viewing geometries.

\begin{figure*}
\begin{center}

\includegraphics[]{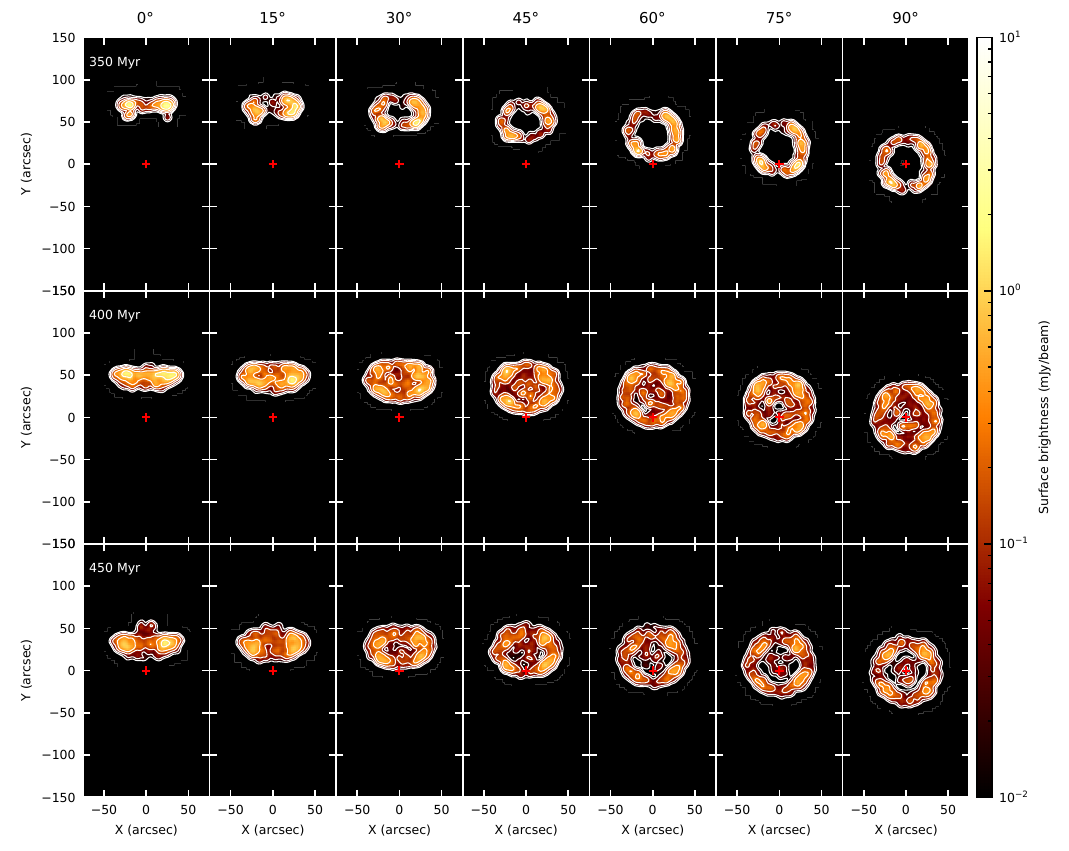}

\caption{Re-energized remnant radio lobes for different observing geometries. Lobes are inflated by a $10^{38}$\,W jet, active for 50\,Myr in a cluster environment; then evolve buoyantly until they are impacted by a plane-parallel normal shock, travelling at 3\,000 km\,s$^{-1}$ in the negative z-direction. 450 Myr snapshot is shown at different viewing angles. This plot is for the same simulation as Figure~\ref{fig:Q38-t50-shock-normal_timeSeries}, but for a larger number of observing geometries.}
\label{fig:Q38-t50-shock-normal_timeSeries_moreAngles}
\end{center}
\end{figure*}

\begin{figure*}
\begin{center}

\includegraphics[]{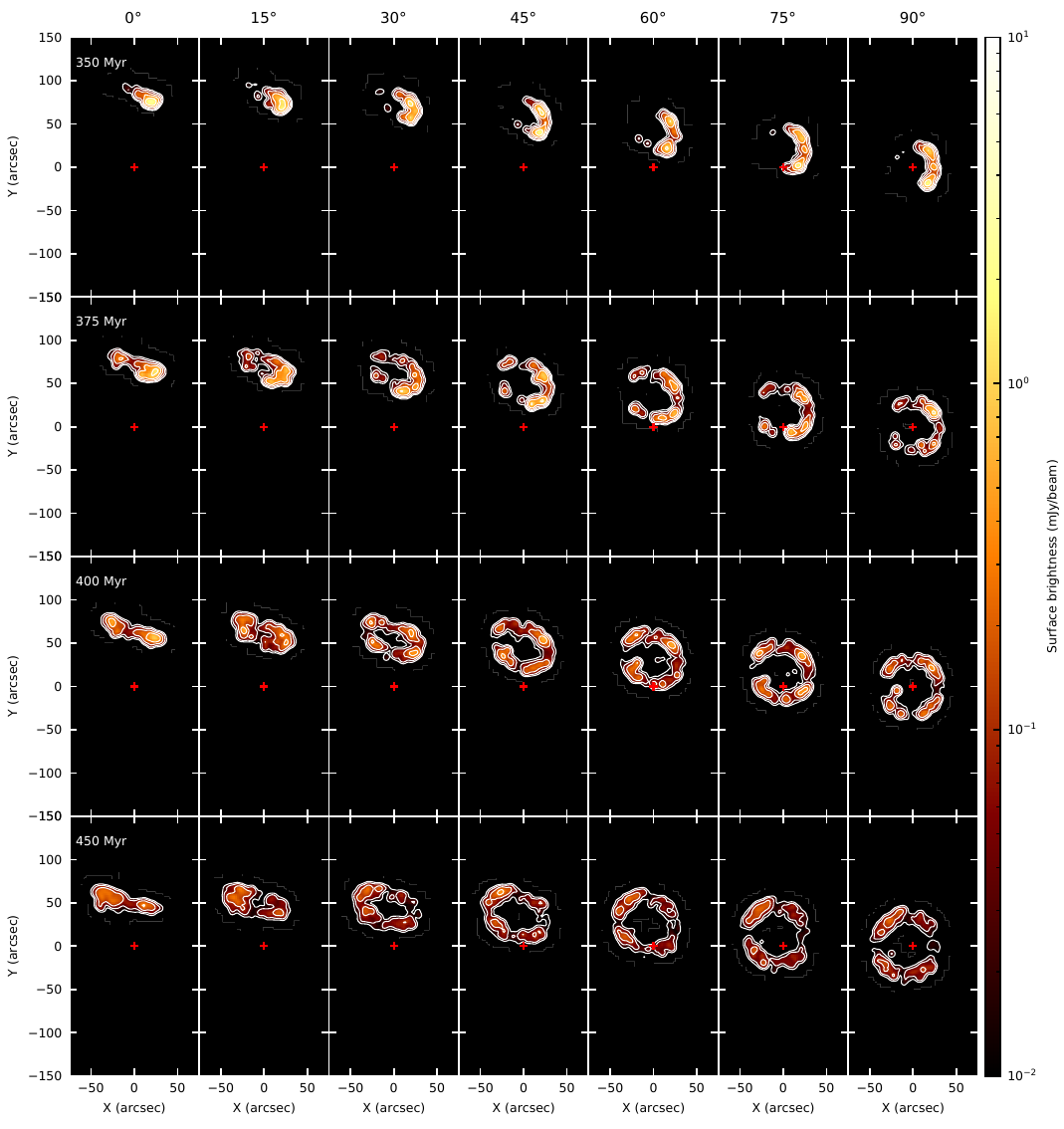}

\caption{Same as Figure~\ref{fig:Q38-t50-shock-normal_timeSeries_moreAngles} but for a shock at 20 degrees to the normal.}
\label{fig:Q38-t50-shock-20deg_timeSeries_moreAngles}
\end{center}
\end{figure*}

\begin{figure*}
\begin{center}

\includegraphics[]{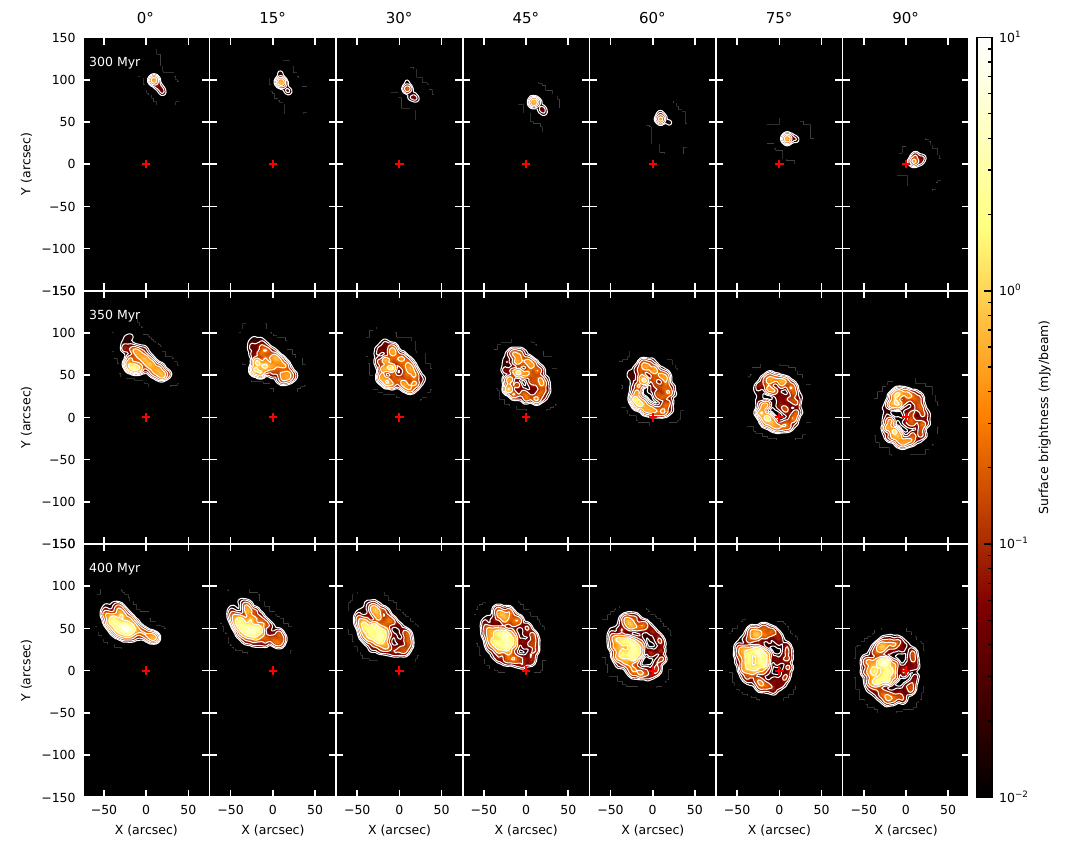}

\caption{Same as Figure~\ref{fig:Q38-t50-shock-normal_timeSeries_moreAngles} but for a shock at 45 degrees to the normal.}
\label{fig:Q38-t50-shock-45deg_timeSeries_moreAngles}
\end{center}
\end{figure*}

\section{Calculation of synchrotron emission}
\label{sec:AppendixRadioEmission}

\subsection{Particle acceleration at shocks}
\label{sec:particleAcceleration}

We use the Lagrangian passive tracer particle module of PLUTO 4.3 \citep{VaidyaEA2018}, and detect shocks following the flagging scheme described by \citet[Appendix B][]{MignoneEA2012}. A zone is flagged as shocked if the divergence of the velocity is negative, $\nabla \cdot {\bf v} <0$, and the local pressure gradient exceeds a threshold $\epsilon_p = p_2/p_1 - 1$, where the subscripts ``1'' and ``2'' refer to pre- and post-shock quantities, respectively. The relationship between shock threshold, Mach number, slope of the power-law distribution of Diffusive Shock Accelerated electrons \citep{BlandfordOstriker1978, Drury1983}, and the injection spectral index, is:
\begin{eqnarray}
\epsilon_p = \frac{p_2}{p_1}-1 = \frac{5(\mathcal{M}^2-1)}{4} \nonumber \\
s = \frac{2(\mathcal{M}^2+1)}{\mathcal{M}^2-1} = \frac{5+2 \epsilon_p}{\epsilon_p}\\
\alpha_{\rm inj}=\frac{s-1}{2} = \frac{5+\epsilon_p}{2 \epsilon_p} \nonumber
\end{eqnarray}
Here, $s$ the power-law index of energy injection $N(E) \propto E^{-s}$, and $\alpha_{\rm inj}$ the injection spectral index for the DSA electrons. For a strong shock ($\mathcal{M} \gg 1$), we recover the canonical result of $s=2$, $\alpha_{\rm inj}=0.5$.

In our simulations, we follow the method of Yates-Jones et al. (2022) to track three pressure discontinuities linearly separated in log-space, $\epsilon_p = [0.5, 1.6, 5]$. These correspond to Mach numbers $\mathcal{M}=[1.1, 1.5, 2.2]$, energy injection indices $s=[12, 5.2, 3]$, and spectral indices at injection $\alpha_{\rm inj} = [5.5, 2.1, 1]$.

In the active jet phase, very strong shocks are found at jet recollimation shocks, and especially hotspots. We set $s_{\rm active}=2.2$, corresponding to an injection spectral index  $\alpha_{\rm inj, active}=0.6$ typically observed in radio galaxies; we note that the exact value of this parameter is not important for the results presented in this paper.

\subsection{Synchrotron emissivity}
\label{sec:PRAiSE}

Synchrotron emissivity is calculated following the approach described in Yates-Jones et al (2022); here we recap key aspects of our approach, and refer the interested reader to that paper for full details.

Electrons are assumed to radiate at the the critical frequency
\begin{equation}
\nu_c = \gamma^2 \nu_{\rm L}
\label{eqn:nu_crit}
\end{equation}
where $\nu_{\rm L}$ is the Larmor frequency \citep[e.g.][]{Longair2011}. The magnetic field strength therefore sets the Lorentz factor of electrons radiating at the observed frequency $\nu$. The rate at which relativistic electrons are injected into the cocoon is set by the jet kinetic power. These electrons are accelerated at strong shocks; Diffusive Shock Acceleration theory predicts a power-law distribution in the energy, and hence Lorentz factor, of electrons at the time of acceleration, with number density
\begin{equation}
n(\gamma)\,d\gamma = n_0 \gamma^{-s}\, d\gamma
\label{eqn:n0}
\end{equation}
The electron energy density subsequently evolves with time due to a combination of adiabatic, synchrotron and Inverse Compton losses; and any re-acceleration at subsequent shocks as described in Section~\ref{sec:particleAcceleration}.

Each Lagrangian particle in the PLUTO simulations represents an ensemble of electrons. For each PLUTO particle, the time of last shock $t_{\rm acc}$ is updated at each simulation snapshot, i.e. every 0.1\,Myr (Section~\ref{sec:simulations}). This time of last electron acceleration is tracked separately for each shock threshold; in our analysis below we focus on $\epsilon_p=5$. Starting with the Lorentz factor required for the electron to radiate at the observed frequency (Equation~\ref{eqn:nu_crit}), we iterate backwards in time to infer the (higher) Lorentz factor $\gamma_{\rm acc}$ at injection time $t_{\rm acc}$ using the recursion relation (see Equation 2 of \citet{YatesJonesEA2022}),

\begin{equation}
\gamma_n = \frac{\gamma_{n-1} \, t_n^{ a_p(t_{n-1},t_n) / 3 \Gamma_{\rm C}} }{t_{n-1}^{ a_p(t_{n-1},t_n) / 3 \Gamma_{\rm C}} - a_2(t_{n-1},t_n) \gamma_{n-1} } 
\label{eqn:t_iterate}
\end{equation}
where $a_p (t_{n-1},t_n) = \log (p_n / p_{n-1}) / \log(t_n/t_{n-1})$ describes the instantaneous pressure evolution, and the radiative loss term $a_2(t_{n-1},t_n)$ depends on the local magnetic field strength (for synchrotron losses) and the Cosmic Microwave Background energy density (for Inverse Compton CMB losses). Electron populations which have suffered severe losses (e.g. in regions of high magnetic field strength and/or accelerated sufficiently long ago) will require very high injection Lorentz factors; such electrons are in the power-law tail of the DSA energy distribution, and will therefore contribute little to the integrated emissivity.

Two further quantities are required to describe the synchrotron emissivity. Simulations presented in this paper are purely hydrodynamic; we therefore need to assume a mapping between the lobe magnetic field strength and a hydrodynamic quantity. Following an established approach \citep{KaiserEA1997, TurnerShabala2015, Hardcastle2018a}, we write
\begin{equation}
p=(\Gamma_{\rm C} - 1) (u_e + u_B + u_T)
\label{ref:p_u_mapping}
\end{equation}
where $u_e$ is the energy density in non-thermal (radiating) particles, $u_B = B^2 / 2\mu_0$ the magnetic field energy density, and $u_T$ the thermal pressure in the cocoon. For FR-II jets considered in this work, constraints from X-ray and radio observations \citep{CrostonEA2018} suggest that there are very few, if any, thermal particles in the cocoon; we therefore set $u_T=0$. We adopt $\eta \equiv u_B / p = 0.03$, which yields lobe magnetic fields at the $\sim 10 \mu$G level, consistent with observations \citep{InesonEA2017}.

In the paradigm examined in this paper, revived lobe emission represents synchrotron radiation from re-accelerated radio galaxy lobes. These are the ``radio-gischt'' of \citet{EnsslinEA1998, HoeftBrueggen2007, IapichinoBrueggen2012}, in which the seed cosmic ray electrons are provided by the now-switched-off jet -- as opposed to compressed fossil plasma which results in ``radio ghosts'' \citep{EnsslinGopalKrishna2001, EnsslinBrueggen2002, KempnerEA2004}. Revived lobes have typical spectral indices $\alpha \sim 1$, consistent with our $\epsilon=5$ ($s=3$) threshold, and we adopt this value here. We take $\eta=0.002$ for the remnant lobes, which yields magnetic fields at the several $\mu$G level, consistent with observations \citep{NakazawaEA2009, FinoguenovEA2010, StroeEA2014}. We note that $\eta$ parametrises important MHD effects not captured in our hydrodynamic simulations, including dynamical amplification of magnetic fields.

The total emissivity at observed frequency $\nu$ is calculated by combining adiabatic and radiative losses (details in \citealt{YatesJonesEA2022}). The number density normalization factor $n_0$ in Equation~\ref{eqn:n0} sets the normalization of emissivity; it is constrained by the total energy injected by the pair of jets, each with kinetic power $Q_{\rm jet}$,

\begin{eqnarray}
2 Q_{\rm j} t_{\rm active} & = & \int_V u_e \, dV \\
& = & \int_V dV\, n_0 m_e c^2 \int_{\gamma_{\rm min}}^{\gamma_{\rm max}} (\gamma-1) \gamma^{-s} \, d\gamma \nonumber \\
& \approx & \int_V dV\, n_0 m_e c^2 \left( \frac{\gamma_{\rm min}^{-(s-1)}}{s-1} \right) \left[ \left( \frac{s-1}{s-2} \right) \gamma_{\rm min} - 1 \right] \nonumber
\end{eqnarray}
where we have assumed electron-positron composition characteristic of FR-II jets \citep{Hardcastle2018a}, and ${\gamma_{\rm min}}$ and ${\gamma_{\rm max}}$ are the Lorentz factors corresponding to low and high energy cutoffs. The integral is insensitive to the choice of $\gamma_{\rm max}$ for typical values of $s>2$ and we make the usual assumption $\gamma_{\rm max} \rightarrow \infty$ \citep{KaiserEA1997, TurnerShabala2015}, yielding the last approximate equality.

The final parameter required to calculate emissivity is the low-energy cutoff for the radiating particles, $\gamma_{\rm min}$. Calculated emissivities depend sensitively on this parameter: for spectra steeper than $s=2$, the majority of electrons will have Lorentz factors just above $\gamma_{\rm min}$. Observations of hotspots in powerful radio sources, similar to those simulated here, show $\gamma_{\rm min} \sim 500$ (e.g. in Cygnus A, \citet{CarilliEA1991, StawarzEA2007, McKeanEA2016} or slightly higher \citet{GodfreyEA2009}). We adopt $\gamma_{\rm min}=500$ for our radio galaxy modelling, noting that this value produces size-luminosity tracks consistent with observations of real radio sources \citep{TurnerEA2018a, TurnerEA2018b, YatesJonesEA2022}.

\subsection{Conservation of particle number density}

Because we simulate the full duty cycle of jet activity, we are able to constrain the total number of emitting particles $N_{\rm rad}$, evaluated at the time when the jet switches off.

The total number of radiating particles is
\begin{equation}
N_{\rm rad} = \frac{2 Q_{\rm jet} t_{\rm active} (s-2)}{m_e c^2 \gamma_{\rm min} }
\label{eqn:Nrad}
\end{equation}

We ensure that this quantity is conserved in the remnant phase of lobe evolution.

\end{document}